\documentclass[11pt,a4paper]{article}

\title{JDOI Variance Reduction Method and the Pricing \\ of American-Style Options}
\author{\sc \Large Johan Auster\footnote{Email: johan.auster@math.ku.dk} \hspace{1em}  Ludovic Mathys\footnote{Email: ludovic.mathys@bf.uzh.ch} \hspace{1em} Fabio Maeder\footnote{Email: f.maeder@math.ethz.ch} \vspace{1.0em} \\
{\it Department of Mathematical Sciences, University of Copenhagen, Denmark.} \vspace{0.2em} \\
       {\it Department of Banking and Finance, University of Zurich, Switzerland.} \vspace{0.2em} \\
			{\it Department of Mathematics, ETH Zurich, Switzerland.} }
\date{}
\providecommand{\keywords}[1]{\textbf{Keywords:} #1}
\providecommand{\mscclass}[2]{\textbf{MSC (2010) Classification:} #1}
\providecommand{\jelclass}[3]{\textbf{JEL Classification:} #1}

\usepackage[utf8]{inputenc}
\usepackage[left=.5in, right=.5in, top=1in, bottom=1in]{geometry}
\usepackage[american]{babel}
\usepackage{parskip}
\usepackage{float}
\usepackage[labelfont=bf]{caption}

\usepackage{glossaries}
\newacronym{doi}{DOI}{Diffusion Operator Integral}
\newacronym{lsmc}{LSMC}{Least-Squares Monte Carlo}
\newacronym{mc}{MC}{Monte Carlo}
\newacronym{gpr}{GPR}{Gaussian Process Regression}
\newacronym{sde}{SDE}{Stochastic Differential Equation}
\newacronym{fd}{FD}{Finite Difference}

\usepackage{tabularx}
\usepackage{booktabs}

\usepackage{lipsum}

\usepackage{amsfonts}
\usepackage{amssymb}
\usepackage{amsmath}
\usepackage{dsfont}
\usepackage{mathtools}

\numberwithin{equation}{section}

\usepackage{subfiles}
\usepackage{xr}

\usepackage[
backend=bibtex,
style=alphabetic,
citestyle=trad-alpha
]{biblatex}

\usepackage{latexsym}
\usepackage{graphicx}
\usepackage{mathrsfs}
\usepackage{amsthm}
\usepackage{mdframed}
\usepackage{bbm}
\usepackage{color}
\usepackage{capt-of}
\usepackage{xcolor}

\begin{document}

%%%   TITLE PAGE  -- START  %%%
\maketitle

\thispagestyle{empty}
% The abstract

\begin{abstract}
\noindent The present article revisits the Diffusion Operator Integral (DOI) variance reduction technique originally proposed in \cite{HP02} and extends its theoretical concept to the pricing of American-style options under (time-homogeneous) Lévy stochastic differential equations. The resulting Jump Diffusion Operator Integral (JDOI) method can be combined with numerous Monte Carlo based stopping-time algorithms, including the ubiquitous least-squares Monte Carlo (LSMC) algorithm of Longstaff and Schwartz (cf.~\cite{Ce96}, \cite{LS01}). We exemplify the usefulness of our theoretical derivations under a concrete, though very general jump-diffusion stochastic volatility dynamics and test the resulting LSMC based version of the JDOI method. The results provide evidence of a strong variance reduction when compared with a simple application of the LSMC algorithm and proves that applying our technique on top of Monte Carlo based pricing schemes provides a powerful way to speed-up these methods.
\end{abstract}
$\;$ \vspace{2em} \\
\noindent \keywords{American Options, Lévy Models, Stochastic Volatility, Variance Reduction, Monte Carlo Methods.} \vspace{0.5em} \\
% Write down at least 3 Keywords
\noindent \mscclass{91-08, 91B25, 91B70, 91G20, 91G60, 91G80.}{} \vspace{0.5em}\\
\noindent \jelclass{C32, C63, G12, G13.} \vspace{-0.5em} \\

%%%   TITLE PAGE  -- END  %%%

% abstract
% \subfile{chapters/abstract}

% introduction

\section{Introduction}
Even after more than five decades of academic research, the pricing of American-style derivatives is still a well studied and challenging problem. Other than their European counterpart, these derivatives are characterized by an early-exercise feature that substantially complicates their structure. In particular, the valuation of American-style derivatives has a direct link to certain types of free-boundary problems, the (analytical) solution of which is only known in a few very special cases. Nevertheless, several numerical techniques have been developed over the years, ranging from hybrid methods that combine analytical and numerical techniques and often lead to very efficient algorithms (cf.~\cite{BW87}, \cite{Ki90}, \cite{Ba91}, \cite{CS14}, \cite{Ma20}), to P(I)DE techiques (cf.~\cite{CV05}, \cite{ZDC07}), tree-based methods (cf.~\cite{BG97}, \cite{BPP03}, \cite{JO12}), as well as Monte Carlo based algorithms (cf.~\cite{Ce96}, \cite{LS01}, \cite{CLP02}, \cite{Eg05}, \cite{EKT07}). In particular the latter stream of methods has gained high relevance in practice, since Monte Carlo based algorithms are usually not tied to particular valuation problems but these can be rather applied to any types of optimal stopping problems. Additionally, the recent advances in machine learning has incentivized research in this particular area and allowed for extensions of these techniques to high-dimensional problems (cf.~\cite{KKT10}, \cite{BCJ19}, \cite{BCJ20}, \cite{GMZ20}, \cite{RW20}). The present article follows this stream of the literature and provides a variance-reduction technique that can be applied on top of numerous Monte Carlo based algorithms, therefore providing a powerful way to speed up these methods. \vspace{1em} \\
\noindent Our paper has several contributions. On the theoretical side, we revisit the Diffusion Operator Integral (DOI) variance reduction technique originally proposed in \cite{HP02} and extend this concept to the pricing of American-style options under (time-homogeneous) Lévy stochastic differential equations (SDEs). In accordance with the naming in the original article (cf.~\cite{HP02}), we refer to the resulting variance reduction method as the Jump Diffusion Operator Integral (JDOI) technique.  Our extension preserves all the benefits of the original DOI method (cf.~\cite{HP02}, \cite{HP14}), while allowing for a great flexibility in the choice of the underlying dynamics. In particular, our method can be adopted to substantially speed up numerous Monte Carlo based (pricing) algorithms. Additionally, while several articles (cf.~\cite{HP02}, \cite{HP14}, \cite{CK18}, \cite{CKD19}) focused on DOI-based methods within pure diffusion settings, our Lévy framework has the advantage to encompass the most frequent dynamics encountered in financial modeling, ranging from single Lévy models to multifactor Lévy stochastic volatility models. Finally, although our theoretical exposition is tied to the problem of pricing American-style derivatives, we emphasize that the very same ideas similarly allow to deal with other types of options, e.g.~with the simpler European-style contracts that were treated as part of applications in \cite{HP02}, \cite{CK18}, and \cite{CKD19}. In fact, to relate our main (American-style) results with the available DOI literature -- that exclusively treats European-style options -- we will briefly review few results for European-style options in our numerical analysis of Section~\ref{SEC_numerical_results}. \vspace{1em} \\
\noindent On the practical side, the usefulness of our theoretical results is exemplied in a concrete, though very general jump-diffusion model that combines the multifactor extension of Grasselli's 4/2 stochastic volatility model (cf.~\cite{Gr17}) with mixed-exponentially distributed jumps (cf.~\cite{CK11}). We call this model specification the Heston$\oplus$3/2$\oplus$Jumps (H3/2J) market model and note that it has the particularity to englobe several of the most important dynamics used in option pricing theory. On the pure diffusion side, the model can be viewed as superposition of two stochastic volatility dynamics, one specified by the Heston dynamics (cf.~\cite{He93}) and the other following a 3/2 model (cf.~\cite{He97}, \cite{Pl97}). Therefore, the resulting diffusion generates an implied volatility surface that can be described by factors with different characteristics. On the pure jump side, relying on mixed-exponential distributions allows to strengthen the model's (analytical) tractability while keeping its generality. Indeed, it is well-known that mixed-exponential distributions are dense -- in the sense of weak convergence -- in the class of all distributions, and that they, therefore, offer the possibility to approximate any jump distribution (cf.~\cite{BH86}, \cite{CK11}, \cite{LV20}, \cite{FMV20}, \cite{FM20}). We provide an extensive discussion of the JDOI method for both American standard and barrier options in this model and extensively test its least-squares Monte Carlo (LSMC) version, i.e.~the version that combines the JDOI method with the LSMC (stopping-time) algorithm of Longstaff and Schwartz (cf.~\cite{Ce96}, \cite{LS01}). The results provide evidence of strong variance reduction when compared with a simple application of the underlying LSMC algorithm and proves that applying the JDOI method on top of Monte Carlo based algorithms provides a powerful way to accelerate these methods. Lastly, we note that our choice in favor of the LSMC algorithm is due to its ubiquity in the financial industry. However, we emphasize that other JDOI versions could be defined, in particular GPR-JDOI and Deep-JDOI versions that combine Gaussian process regression based stopping-time algorithms (cf.~\cite{GMZ20}) and (deep) neural network based stopping-time algorithms (cf.~\cite{BCJ19}, \cite{BCJ20}), respectively, with our JDOI variance reduction technique. Investigating these types of algorithms could be part of future research. \vspace{1em} \\
\noindent The remaining of the paper is structured as follows. In Section~\ref{SEC_Method}, we introduce our JDOI method for (time-homogeneous) Lévy SDEs as well as the notation used in the rest of the paper. This section also discusses conditions under which a substantial variance-reduction may be achieved. The presentation therein is tied to the pricing of American-style options, although the general ideas underlying our method could be similarly extended to other types of options. Section~\ref{SEC_Applications} deals with applications of the JDOI technique under the H3/2J model dynamics. Here, a natural approximate market for this model is introduced and theoretical derivations of the JDOI method for American standard and barrier options are provided. Finally, the theoretical considerations of Sections~\ref{SEC_Method}-\ref{SEC_Applications} are tested in Section~\ref{SEC_numerical_results} and the paper concludes with Section~\ref{SEC_Conclu}. Complementary results are presented as part of the appendices (Appendix A~and~B).

% doi methods for american options under levy sdes

\section{JDOI Method and American-Type Option Pricing Problems}
\label{SEC_Method}
In this section, we extend the \gls{doi} method presented in \cite{HP02} to the case of (time-homogeneous) Lévy \glspl{sde}. Our approach focuses on American-style options, since these are the type of options we ultimately want to price. However, we note that the very same approach similarly allows to derive a generalization of the \gls{doi} method for other types of options, in particular for the simple European-style contracts. Our notation is inspired by the two great works on stochastic control of jump-diffusions and on optimal stopping in \cite{OS07} and \cite{PS06}, respectively.

\subsection{General Setting and Pricing of American-Type Options\label{GeneralSetting}}
We work on a filtered probability space $\left(\Omega, \mathcal{F}, \mathbf{F}, \mathbb{Q} \right)$ $-$ a (chosen) risk-neutral probability space $-$ satisfying the usual conditions and consider, for $T >0$, a ($d+1$)-dimensional financial market consisting of a savings account $(X_{t}^{0})_{t \in [0,T]}$ and risky assets whose dynamics are fully described by a $d$-dimensional stochastic process, $ (X_{t})_{t \in [0,T]} = \big \{ X_{t} = (X_{t}^{1}, \ldots, X_{t}^{d} \big): \, t \in [0,T] \big\}$, representing asset prices and/or state variables. We assume that the dynamics of the savings account is described, for a function $r: \mathbb{R}^{d} \rightarrow [0,\infty)$, by
\begin{align}
d X_{t}^{0}  = r(X_{t})  X_{t}^{0} \, dt, \hspace{1.5em} t \in [0,T], 
\label{EqDynRN}
\end{align}
with $X_{0}^{0}  = 1$, and that each of the remaining entries $(X_{t}^{i})_{t \in [0,T]}$, $i\in \{1, \ldots, d\}$, evolves according to a (time-homogeneous) Lévy SDE\footnote{Time-homogenenous SDEs are usually referred to as jump-diffusions or Lévy diffusions (cf.~\cite{OS07}).}
\begin{equation}
dX_{t}^{i} = b^{i}(X_{t}) \,dt + \sum \limits_{j=1}^{m} \sigma^{ij}(X_{t}) \, dW_{t}^{j} + \sum \limits_{k=1}^{\ell} \int \limits_{\mathbb{R}} \gamma^{ik}(X_{t-},z_{k}) \, \tilde{N}_{k}(dt,dz_{k}), \hspace{1.5em} t \in [0,T],
\label{EqDyn}
\end{equation}
where $X_{0}^{i} = x_{i} \in \mathbb{R}$, $(W_{t})_{t \in [0,T]} = \big\{ W_{t} = (W_{t}^{1}, \ldots, W_{t}^{m}): \, t \in [0,T] \big\}$ is an $m$-dimensional Brownian motion and $\tilde{N}(dt,dz) = \big(\tilde{N}_{1}(dt,dz_{1}), \ldots, \tilde{N}_{\ell}(dt,dz_{\ell}) \big)$ refers to an $\ell$-dimensional vector of independent compensated Poisson random measures. Here, for any $i\in \{1, \ldots, d\}$ the coefficients $b^{i}: \mathbb{R}^{d} \rightarrow \mathbb{R}$, $\sigma^{ij}: \mathbb{R}^{d} \rightarrow \mathbb{R}$, $j \in \{1, \ldots, m \}$, and $\gamma^{ik}: \mathbb{R}^d \times \mathbb{R} \rightarrow \mathbb{R}$, $k\in \{1,\ldots,\ell \}$, are assumed to satisfy appropriate conditions such that (\ref{EqDyn}) admits a unique strong solution.\footnote{In particular, linear growth and Lipschitz conditions, as outlined in Theorem~1.19 in \cite{OS07}, are sufficient to ensure the existence of a strong solution to (\ref{EqDyn}).}~In this case, the obtained solution is known to be strongly Markovian. Additionally, we assume that the resulting market is arbitrage-free in the sense that discounted asset dynamics are true martingales under the pricing measure $\mathbb{Q}$. Finally, we will always work under the filtration generated by $W$ and $\tilde{N}$, i.e.~$\mathbf{F} := (\mathcal{F}_{t})_{t \in [0,T]}$ will always refer to the augmented natural filtration of $(W_t)_{t \in [0,T]}$ and $\big\{ \tilde{N}_{k}((0,t],A): \, t \in [0,T], \, A \in \mathcal{B}\setminus \{0\} \big\}$, $k \in \{1,\ldots,\ell \}$. \vspace{1em} \\
Given a (measurable) payoff function $G: [0,T] \times \mathbb{R}^{d} \rightarrow \mathbb{R}$ satisfying, for any $(t,x) \in [0,T] \times \mathbb{R}^{d}$, the condition\footnote{As noted in \cite{PS06}, this condition is naturally satisfied in many option pricing problems.}
\begin{equation}
\mathbb{E}_{t,x}^{\mathbb{Q}} \bigg[ \, \sup \limits_{0 \leq u \leq T-t} \Big| \frac{X_{t}^{0} \, G(t+u,X_{u})}{X_{t+u}^{0}} \Big | \, \bigg] < \infty,
\label{IntCond}
\end{equation}
we consider the following (finite-horizon) optimal stopping problem
\begin{equation}
V_{A}^{X}(t,x) :=  \sup \limits_{ \tau \in \mathfrak{T}_{[0,T-t]} } \mathbb{E}_{t,x}^{\mathbb{Q}} \left[ \frac{X_{t}^{0} G(t+\tau,X_{\tau})}{X_{t+\tau}^{0}}  \right] = \sup \limits_{ \tau \in \mathfrak{T}_{[0,T-t]} } \mathbb{E}_{t,x}^{\mathbb{Q}} \left[ e^{-\int_{t}^{t+\tau} r(X_{s})\, ds} \, G(t+\tau, X_{\tau}) \right] ,
\label{ProbOS}
\end{equation}
where $\mathbb{E}_{t,x}^{\mathbb{Q}}[ \cdot ]$ denotes expectation under the measure $\mathbb{Q}_{t,x}$ having initial value $Z_{0} =(t,x)$, for the (strong Markov) process $(Z_{t})_{t \in [0,T]} := \big\{ (t,X_{t}): \, t \in [0,T] \big \}$, and $\mathfrak{T}_{[0,T-t]}$ denotes the set of stopping times that take values in the interval $[0,T-t]$. Under mild additional conditions on the functions $V_{A}^{X}(\cdot)$ and $G(\cdot)$ an optimal stopping time to Problem (\ref{ProbOS}) can be derived. Indeed, if $V_{A}^{X}(\cdot)$ and $G(\cdot)$ are lower and upper semi-continuous respectively,\footnote{We emphasize that these are very natural conditions that are often satisfied in practice. In particular, this holds true in our applications of Section~\ref{SEC_Applications} (cf.~Remark 2.10.~in \cite{PS06}).}~well-known optimal stopping results (cf.~Corollary 2.9.~in \cite{PS06}) imply that, for any $t \in [0,T]$, the following first-entry time
\begin{equation}
\tau_{\mathcal{D}_{s}} := \inf \{ 0 \leq u \leq T-t : \, (t+u,X_{u}) \in \mathcal{D}_{s} \},
\label{OptimalStopping}
\end{equation}
where continuation and stopping regions, $\mathcal{D}_{c}$ and $\mathcal{D}_{s}$, are defined via
\begin{align}
\mathcal{D}_{c} &:= \{ (t,x) \in [0,T] \times \mathbb{R}^{d}: \, V_{A}^{X}(t,x) > G(t,x) \}, \\
\mathcal{D}_{s} &:= \{ (t,x) \in [0,T] \times \mathbb{R}^{d}: \, V_{A}^{X}(t,x) = G(t,x) \},
\end{align}
is optimal in Problem (\ref{ProbOS}). Hence, we have that
\begin{equation}
V_{A}^{X}(t,x) =  \mathbb{E}_{t,x}^{\mathbb{Q}} \left[ e^{ -\int_{t}^{t+\tau_{\mathcal{D}_{s}}} r(X_{s}) \, ds} \, G\big(t+\tau_{\mathcal{D}_{s}}, X_{\tau_{\mathcal{D}_{s}}}\big) \right].
\label{REPREimp}
\end{equation}
In particular, the latter representation allows to further characterize the value function $V_{A}^{X}(\cdot)$ in terms of a Cauchy-type problem. Indeed, if one notes that, for sufficiently smooth functions $V:[0,T] \times \mathbb{R}^{d} \rightarrow \mathbb{R}$, the infinitesimal generator associated to $(X_t)_{t \in [0,T]}$ is given by
\begin{align}
\mathcal{A}_{X}V(t,x) & := \lim \limits_{ h \downarrow 0} \frac{\mathbb{E}_{x}^{\mathbb{Q}} \left[ V(t,X_{h}) - V(t,x) \right]}{h} \nonumber \\
& = \sum \limits_{i=1}^{d} b^{i}(x) \, \partial_{x_{i}} V(t,x)  + \frac{1}{2} \sum \limits_{i=1}^{d} \sum \limits_{j=1}^{d} \left( \sigma(x) \sigma(x)^{\intercal} \right)_{ij} \, \partial_{x_{i}} \partial_{x_{j}} V(t,x)  \nonumber \\
& \hspace{2.5em} + \sum \limits_{k=1}^{\ell} \, \int \limits_{\mathbb{R}} \Big( V\big(t,x + \gamma^{\star,k}(x,z_{k}) \big) - V(t,x) - \sum \limits_{i=1}^{d} \gamma^{ik}(x,z_{k}) \, \partial_{x_{i}} V(t,x) \Big) \Pi_k(z_k),
\label{INFIgen}
\end{align}
where $\sigma(\cdot)$ refers to the $d \times m$ matrix defined by $\sigma(x):= \left[\sigma^{ij}(x)\right]_{ij}$ and $\gamma^{\star,k}(\cdot)$ is the $k$-th column of the $d \times \ell$ matrix $\gamma(x,z) := \left[ \gamma^{ik}(x,z_{k}) \right]_{ik}$, then standard (strong) Markovian arguments show that $V_{A}^{X}(\cdot)$ solves the following Cauchy-type problem: 
\begin{align}
\partial_{t} V_{A}^{X}(t,x) + \mathcal{A}_{X} V_{A}^{X}(t,x) & = r(x) V_{A}^{X}(t,x), \hspace{1.5em} \mbox{on} \; \, \mathcal{D}_{c}, \label{PRO1}\\
V_{A}^{X}(t,x)  & = G(t,x), \hspace{4em} \mbox{on} \; \, \mathcal{D}_{s}. \label{PRO2}
\end{align}
Together with Equation (\ref{REPREimp}), Characterizations (\ref{PRO1}), (\ref{PRO2}) form the basis of many methods of solution used in practice. However, recovering the value function $V_{A}^{X}(\cdot)$ analytically is a challenging task so that derivations are usually done numerically, with several streams arising in practice -- in particular:
\begin{itemize} \setlength \itemsep{-0.2em}
\item Hybrid methods, combining analytical and numerical techniques (cf.~\cite{BW87}, \cite{Ki90}, \cite{Ba91}, \cite{CS14}, \cite{Ma20}).
\item Numerical methods for P(I)DEs, such as finite difference and finite elements methods as well as extensions thereof, (cf.~\cite{CV05}, \cite{ZDC07}).
\item Tree-based algorithms (cf.~\cite{BG97}, \cite{BPP03}, \cite{JO12}).
\item Monte Carlo methods, such as the Least-Squares Monte Carlo (LSMC) approach of Longstaff and Schwartz (cf.~\cite{Ce96}, \cite{LS01}) and its extensions (cf.~\cite{CLP02}, \cite{Eg05}, \cite{EKT07}).
\end{itemize}
The present work follows the line of Monte Carlo methods and provides a variance-reduction technique that can be combined with numerous Monte Carlo based algorithms, therefore providing a powerful way to speed up these methods. This is discussed in the next section. \vspace{1em} \\
\noindent \underline{\bf Remark 1.} \vspace{0.2em} \\
We already emphasize that our general framework is able to deal with a wide range of American-style problems. In fact, although the above description is at first tied to American-style options with exercise payoffs depending on the process's current value only, switching to Markov processes that are killed when reaching a certain boundary often allows for a straight extension of our framework to problems with exercise payoffs depending on the full path history. This holds in particular true when dealing with various types of occupation time derivatives.\footnote{cf.~\cite{FM20} for an embedding of the full class of (geometric) double-barrier step options into simple option pricing problems using killed Lévy processes.}~Similarly, when pricing double-barrier contracts with $d$ pairs of lower and upper barrier levels $(L_{1}, H_{1}), \ldots, (L_{d}, H_{d})$, having the form
\begin{align}
V_{A}^{X,\mathcal{DB}}(t,x) & := \sup \limits_{ \tau \in \mathfrak{T}_{[0,T-t]} } \mathbb{E}_{t,x}^{\mathbb{Q}} \left[ \frac{X_{t}^{0} G(t+\tau,X_{\tau})}{X_{t+\tau}^{0}} \prod \limits_{i=1}^{d} \mathds{1}_{ \{ L_{i} \leq  X_{\tau}^{i} \leq H_{i} \}}	 \right] \nonumber \\
& = \sup \limits_{ \tau \in \mathfrak{T}_{[0,T-t]} } \mathbb{E}_{t,x}^{\mathbb{Q}} \left[ e^{-\int_{t}^{t+\tau} r(X_{s})\, ds} \, G(t+\tau, X_{\tau}) \prod \limits_{i=1}^{d} \mathds{1}_{ \{ L_{i} \leq  X_{\tau}^{i} \leq H_{i} \}}	 \right] , \label{DBgenEqua}
\end{align}
we can define a new state $\partial$, introduce the set of corresponding cemetery states 
\begin{equation}
\mathcal{S}_{\partial}:= \big \{ x=(x_{1}, \ldots, x_{d}) \in \mathbb{R}^{d}: \, \exists i \in \{1, \ldots, d \}: \, x_{i} =\partial \big \}, 
\end{equation}
\noindent and switch from $(X_{t})_{t \in [0,T]}$ to the process $(X_{t}^{\boldsymbol{\ell}})_{t \in [0,T]}$, defined, with $\boldsymbol{\ell} := (\boldsymbol{\ell}_{1}, \ldots, \boldsymbol{\ell}_{d})$, and $\boldsymbol{\ell}_{i} := (L_{i}, H_{i})$, $i = 1, \ldots , d$, on the domain $\mathcal{D} := [0,T] \times \Big( \bigtimes \limits_{i=1}^{d} \big( [L_{i},H_{i}] \cup \{ \partial \}\big) \Big)$ via 
\begin{equation}
X_{t}^{\boldsymbol{\ell}} := \big(X_{t}^{1,\boldsymbol{\ell}_{1}}, \ldots, X_{t}^{d,\boldsymbol{\ell}_{d}}\big) , \hspace{1.5em} t \in [0,T],
\end{equation}
\noindent where the process components $(X_{t}^{i,\boldsymbol{\ell}_{i}})_{t \in [0,T]}$ are recovered, for $t \in [0,T]$, by means of the relation
\begin{align}
X_{t}^{i,\boldsymbol{\ell}_{i}} :=  X_{t \wedge \tau_{\boldsymbol{\ell}_{i}}}^{i}, \hspace{1.5em} \mbox{if} \;  \, X_{t \wedge \tau_{\boldsymbol{\ell}_{i}}}^{i} \in  [L_{i}, H_{i}] ,\\
 X_{t}^{i,\boldsymbol{\ell}_{i}}  := \partial , \hspace{1.5em} \mbox{otherwise}, \hspace{2.8em} 
\end{align}
\noindent and $\tau_{\boldsymbol{\ell}_{i}} := \inf \left \{ t \geq 0: \, X_{t}^{i} \notin (L_{i}, H_{i}) \right \}$ denotes the first exit time from the interval $(L_{i}, H_{i})$. Then, setting $ G(\cdot, x) \equiv 0$, for $x \in \mathcal{S}_{\partial}$, Equation (\ref{DBgenEqua}) can be re-expressed in the form
\begin{equation}
V_{A}^{X,\mathcal{DB}}(t,x) := \sup \limits_{ \tau \in \mathfrak{T}_{[0,T-t]} } \mathbb{E}_{t,x}^{\mathbb{Q}} \left[ e^{-\int_{t}^{t+\tau} r\left(X_{s}^{\boldsymbol{\ell}}\right)\, ds} \, G\big(t+\tau, X_{\tau}^{\boldsymbol{\ell}}\big)	 \right] ,
\end{equation}
\noindent and using the fact that the newly created process $(X_{t}^{\boldsymbol{\ell}})_{t \in [0,T]}$ behaves exactly like $(X_{t})_{t \in [0,T]}$ for all times $t \leq \tau_{\boldsymbol{\ell}} :=  \tau_{\boldsymbol{\ell}_{1}} \wedge \ldots \wedge \tau_{\boldsymbol{\ell}_{d}}$, one arrives again at Problem (\ref{PRO1})-(\ref{PRO2}). However, it is important to note that the stopping region takes now a slightly different form. In particular, it can be decomposed into two disjoint sets
\begin{equation}
\mathcal{D}_{s} = \mathcal{D}_{s}^{\mathcal{S}_{\partial}} \, \dot \cup \, \mathcal{D}_{s}^{\ast}, \hspace{1.75em} \mbox{with} \hspace{1.75em} \mathcal{D}_{s}^{\mathcal{S}_{\partial}} := \mathcal{D} \cap \mathcal{S}_{\partial} \hspace{1em} \mbox{and} \hspace{1em} \mathcal{D}_{s}^{\ast} := \mathcal{D}_{s} \setminus \mathcal{D}_{s}^{\mathcal{S}_{\partial}},  \label{DecompSet}
\end{equation}
i.e.~$\mathcal{D}_{s}^{\mathcal{S}_{\partial}}$ represents a stale set where the process is already killed and the value of the American-type option remains unchanged $V_{A}^{X,\mathcal{DB}}(\cdot) \equiv 0$. This decomposition will also impact our final JDOI estimate, as described in~Remark~2. \vspace{0.2em} \\
\mbox{ } \hspace{42.5em} \scalebox{0.75}{$\blacklozenge$}
\subsection{Derivation of the Method}
Instead of applying crude Monte Carlo methods to solve Valuation Problem (\ref{REPREimp}) under the market dynamics given in (\ref{EqDynRN}) and (\ref{EqDyn}), it may be beneficial, both from the point of view of computational intensity and pricing precision, to switch to an easier, approximate valuation problem and to subsequently focus on the study of the resulting approximation error. This is the general idea underlying variance-reduction methods as well as several numerical approaches to the pricing of American-style options. Indeed, when dealing with American-style options, a natural approximation is given by the respective European-style contract, whose price is often easier to recover. This subsequently reduces the pricing attempt to the derivation of an early exercise premium that only makes up a small fraction of the full American-style option price (cf.~\cite{BW87}, \cite{Ba91}, \cite{Ba05}, \cite{CS14}, \cite{Ma20}, \cite{FM20}). \vspace{1em} \\
Our general ansatz makes use of all these ideas. However, instead of approximating American-style options with their European counterparts under the same market dynamics, we additionally propose to switch to a (simpler) approximate market. Therefore, instead of relying on the dynamics described in (\ref{EqDynRN}) and (\ref{EqDyn}), we consider another approximate ($d+1$)-dimensional financial market on $\left(\Omega, \mathcal{F}, \mathbf{F}, \mathbb{Q}\right)$ that consists of a savings account $(\bar{X}_{t}^{0})_{t \in [0,T]}$, with $\bar{X}_{0}^{0}=1$ and
\begin{align}
d \bar{X}_{t}^{0}  = r(\bar{X}_{t})  \bar{X}_{t}^{0} \, dt, \hspace{1.5em} t \in [0,T], 
\label{APPROXEqDynRN}
\end{align}
and risky assets whose dynamics are described by a $d$-dimensional stochastic process representing asset prices and/or state variables, $ (\bar{X}_{t})_{t \in [0,T]} = \big \{ \bar{X}_{t} = (\bar{X}_{t}^{1}, \ldots, \bar{X}_{t}^{d} \big): \, t \in [0,T] \big\}$, and that satisfies  
\begin{equation}
d\bar{X}_{t}^{i} = \bar{b}^{i}(\bar{X}_{t}) \,dt + \sum \limits_{j=1}^{m} \bar{\sigma}^{ij}(\bar{X}_{t}) \, dW_{t}^{j} + \sum \limits_{k=1}^{\ell} \int \limits_{\mathbb{R}} \bar{\gamma}^{ik}(\bar{X}_{t-},z_{k}) \, \tilde{N}_{k}(dt,dz_{k}), \hspace{1.5em} t \in [0,T],
\label{APPROXEqDyn}
\end{equation}
with $\bar{X}_{0}^{i} = x_i \in \mathbb{R}$. As earlier, the coefficients $\bar{b}^{i}: \mathbb{R}^{d} \rightarrow \mathbb{R}$, $\bar{\sigma}^{ij}: \mathbb{R}^{d} \rightarrow \mathbb{R}$, $j \in \{1, \ldots, m \}$, and $\bar{\gamma}^{ik}: \mathbb{R}^d \times \mathbb{R} \rightarrow \mathbb{R}$, $k\in \{1,\ldots,\ell \}$, are assumed to satisfy for any $i\in \{1, \ldots, d\}$ appropriate conditions such that (\ref{EqDyn}) admits a unique strong solution and that the resulting discounted asset dynamics are true martingales under the pricing measure $\mathbb{Q}$. \vspace{1em} \\
Under this new approximate market, we consider the European-style option given by 
\begin{equation}
V_{E}^{\bar{X}}(t,x) := \mathbb{E}_{t,x}^{\mathbb{Q}} \left[ \frac{\bar{X}_{t}^{0} G(T,\bar{X}_{T-t})}{\bar{X}_{T}^{0}}  \right] = \mathbb{E}_{t,x}^{\mathbb{Q}} \left[ e^{-\int_{t}^{T} r(\bar{X}_{s}) \, ds } \, G\big(T, \bar{X}_{T-t}\big) \right],
\end{equation}
and assume the following additional integrability condition
\begin{equation}
\mathbb{E}_{t,x}^{\mathbb{Q}} \bigg[ \, \sup \limits_{0 \leq u \leq T-t} \Big| \frac{\bar{X}_{t}^{0} \, G(t+u,\bar{X}_{u})}{\bar{X}_{t+u}^{0}} \Big |^{2} \, \bigg] < \infty, \hspace{1.5em} (t,x) \in [0,T] \times \mathbb{R}^{d}.
\label{IntCond}
\end{equation}
The latter condition is slightly stronger than (\ref{IntCond}) and implies in particular, for any initial value $Z_{0} = (t,x)$, that the discounted price processes $(M_{u}^{E})_{u \in [t,T]}$ and $(M_{u}^{A})_{u \in [t,T]}$, defined for $\bullet \in \{E,A \}$ via
\begin{equation}
M_{u}^{\bullet} := e^{-\int_{t}^{t+u} r(\bar{X}_{s}) \, ds} \, V_{\bullet}^{\bar{X}}(t+u,\bar{X}_{u}), \hspace{1.5em} u \in [t,T],
\label{SquareIntAssumption}
\end{equation}
are square-integrable martingales. Indeed, while the martingale property follows from standard (strong) Markovian arguments (cf.~\cite{PS06}), Doob's maximal inequality (cf.~\cite{RY99}) allows to derive that
\begin{align}
\mathbb{E}_{t,x}^{\mathbb{Q}}\Big[ \, \sup \limits_{t \leq u \leq T} \big|M_{u}^{\bullet} \big|^{2} \, \Big] & \leq 4 \, \mathbb{E}_{t,x}^{\mathbb{Q}} \Big[ \, \big| M_{T}^{\bullet} \big|^2\, \Big] \nonumber \\
& \leq 4 \, \mathbb{E}_{t,x}^{\mathbb{Q}} \Big[ \, \sup \limits_{0 \leq t \leq T} \big| \big(\bar{X}_{t}^{0}\big)^{-1} G(t,\bar{X}_{t}) \big |^{2} \, \Big] < \infty,
\label{Square_INT}
\end{align}
which already shows the square-integrability. Additionally, standard (strong) Markovian arguments imply that the European-style option, $V_{E}^{\bar{X}}(\cdot)$, satisfies the following Cauchy-type problem
\begin{align}
\partial_{t} V_{E}^{\bar{X}}(t,x) + \mathcal{A}_{\bar{X}} V_{E}^{\bar{X}}(t,x) & = r(x) V_{E}^{\bar{X}}(t,x), \hspace{1.5em} \mbox{on} \; \, [0,T) \times \mathbb{R}^{d}, \label{PROEuro1}\\
V_{E}^{\bar{X}}(T,x)  & = G(T,x), \hspace{5em} \mbox{on} \; \,\mathbb{R}^{d}, \label{PROEuro2}
\end{align}
where $\mathcal{A}_{\bar{X}}$ denotes the infinitesimal generator of $\bar{X}$ that is obtained as in (\ref{INFIgen}) while replacing the functions $b^{i}(\cdot)$, $\sigma^{ij}(\cdot)$, $j \in \{1, \ldots,m \}$, and $\gamma^{ik}(\cdot)$, $k\in \{1, \ldots, \ell \}$, for $i \in \{1, \ldots, d \}$ by $\bar{b}^{i}(\cdot)$, $\bar{\sigma}^{ij}(\cdot)$, $j \in \{1, \ldots,m \}$, and $\bar{\gamma}^{ik}(\cdot)$, $k\in \{1, \ldots, \ell \}$.  These results can now be used to derive an unbiased estimator of $V_{A}^{X}(\cdot)$ that is based on integral representations. Indeed, whenever $V_{E}^{\bar{X}}(\cdot)$ has sufficient regularity, e.g.~$V_{E}^{\bar{X}}(\cdot) \in C^{1,2}\left([0,T] \times \mathbb{R}^{d} \right)$, Itô's formula can be combined with Problem (\ref{PROEuro1}), (\ref{PROEuro2}) to derive, for any $t \in [0,T]$, that
\begin{align}
e^{-\int_{t}^{t+\tau_{\mathcal{D}_{s}}} r(X_{s}) \, ds } & \, V_{E}^{\bar{X}}(t+\tau_{\mathcal{D}_{s}},X_{\tau_{\mathcal{D}_{s}}}) = V_{E}^{\bar{X}}(t,x) + \int_{0}^{\tau_{\mathcal{D}_{s}}} e^{-\int_{t}^{t+u} r(X_{s}) \, ds } \left( \mathcal{A}_{X} - \mathcal{A}_{\bar{X}} \right) V_{E}^{\bar{X}}(t+u,X_{u}) \, du  \nonumber \\
& + \sum \limits_{j=1}^{m} \int_{0}^{\tau_{\mathcal{D}_{s}}} \sum \limits_{i=1}^{d} e^{-\int_{t}^{t+u} r(X_{s}) \, ds} \,\sigma^{ij}(X_{u}) \, \partial_{x_{i}} V_{E}^{\bar{X}}(t+u,X_{u}) \,dW_{u}^{j} \nonumber \\
& + \sum \limits_{k=1}^{\ell} \int_{0}^{\tau_{\mathcal{D}_{s}}} \int_{\mathbb{R}}^{} e^{-\int_{t}^{t+u} r(X_{s}) \, ds} \Big[ V_{E}^{\bar{X}}\big(t+u,X_{u-} + \gamma^{\star,k}(X_{u-},z_{k}) \big) - V_{E}^{\bar{X}}(t+u,X_{u-}) \Big] \tilde{N}_{k}(du,dz_{k}).
\label{DERIV_Euro_Part}
\end{align}
However, at the same time, it is easily seen that
\begin{align}
V_{A}^{X}(t,x) &  = \mathbb{E}_{t,x}^{\mathbb{Q}} \Big[ e^{-\int_{t}^{t+\tau_{\mathcal{D}_{s}}} r(X_{s})\, ds } \, V_{A}^{X}\big(t+\tau_{\mathcal{D}_{s}}, X_{\tau_{\mathcal{D}_{s}}}\big) \Big]  \nonumber \\
& = \mathbb{E}_{t,x}^{\mathbb{Q}} \Big[ e^{-\int_{t}^{t+\tau_{\mathcal{D}_{s}}} r(X_{s}) \, ds } \big(  V_{A}^{X}(t+\tau_{\mathcal{D}_{s}}, X_{\tau_{\mathcal{D}_{s}}}) - V_{E}^{\bar{X}}(t+\tau_{\mathcal{D}_{s}}, X_{\tau_{\mathcal{D}_{s}}}) \big)\Big]  \hspace{9em}  \nonumber\\
& \hspace{15em} + \mathbb{E}_{t,x}^{\mathbb{Q}} \Big[ e^{-\int_{t}^{t+\tau_{\mathcal{D}_{s}}} r(X_{s}) \, ds } \, V_{E}^{\bar{X}}\big(t+\tau_{\mathcal{D}_{s}}, X_{\tau_{\mathcal{D}_{s}}}\big) \Big]. 
\end{align}
Therefore, upon assuming that, for any starting values $(t,x) \in [0,T] \times \mathbb{R}^{d}$, we have
\begin{equation}
\mathbb{E}_{t,x}^{\mathbb{Q}} \left[ \, \int_{0}^{T-t} \big| \overline{\mathcal{R}}_{W,X}^{j}(t,u) \big|^{2} \, du \, \right] < \infty, \hspace{2em} \mathbb{E}_{t,x}^{\mathbb{Q}} \left[ \, \int_{0}^{T-t} \int_{\mathbb{R}}^{} \, \big| \overline{\mathcal{R}}_{N,X}^{k}(t,u,z_{k}) \big|^{2} \, \Pi_{k}(dz_{k}) \, du \, \right] < \infty,
\end{equation}
for all $j \in \{1, \ldots, m \}$ and $k \in \{ 1, \ldots, \ell \}$, with
\begin{align}
& \overline{\mathcal{R}}_{W,X}^{j}(t,u)  := \sum \limits_{i=1}^{d} e^{-\int_{t}^{t+u} r(X_{s}) \, ds} \,\sigma^{ij}(X_{u}) \, \partial_{x_{i}} V_{E}^{\bar{X}}(t+u,X_{u}), \hspace{1.5em} j \in \{1, \ldots, m \}, \label{Abbrev1}\\
\overline{\mathcal{R}}_{N,X}^k(t,u,z_{k}&) := e^{-\int_{t}^{t+u} r(X_{s}) \, ds} \Big[ V_{E}^{\bar{X}}\big(t+u,X_{u-} + \gamma^{\star,k}(X_{u-},z_{k}) \big) - V_{E}^{\bar{X}}(t+u,X_{u-}) \Big], \hspace{1.5em} k \in \{1, \ldots, \ell \}, \label{Abbrev2}
\end{align}
one arrives at 
\begin{align}
V_{A}^{X}(t,x) & = V_{E}^{\bar{X}}(t,x) + \mathbb{E}_{t,x}^{\mathbb{Q}} \Big[ e^{-\int_{t}^{t+\tau_{\mathcal{D}_{s}}} r(X_{s}) \, ds } \big(  V_{A}^{X}(t+\tau_{\mathcal{D}_{s}}, X_{\tau_{\mathcal{D}_{s}}}) - V_{E}^{\bar{X}}(t+\tau_{\mathcal{D}_{s}}, X_{\tau_{\mathcal{D}_{s}}}) \big)\Big] \hspace{9em} \nonumber \\
& \hspace{13em} + \mathbb{E}_{t,x}^{\mathbb{Q}} \bigg[ \, \int_{0}^{\tau_{\mathcal{D}_{s}}} e^{-\int_{t}^{t+u} r(X_{s}) \, ds } \left( \mathcal{A}_{X} - \mathcal{A}_{\bar{X}} \right) V_{E}^{\bar{X}}(t+u,X_{u}) \, du \,\bigg].
\end{align}
This finally shows that, under the initial condition $Z_{0} = (t,x)$, the following jump-diffusion operator integral (JDOI) estimator
\begin{align}
\mathcal{Z}_{t,x,\tau_{{D}_{s}}}^{X} & := V_{E}^{\bar{X}}(t,x) + e^{-\int_{t}^{t+\tau_{\mathcal{D}_{s}}} r(X_{s}) \, ds } \big(  V_{A}^{X}(t+\tau_{\mathcal{D}_{s}}, X_{\tau_{\mathcal{D}_{s}}}) - V_{E}^{\bar{X}}(t+\tau_{\mathcal{D}_{s}}, X_{\tau_{\mathcal{D}_{s}}}) \big) \hspace{9em} \nonumber \\
& \hspace{13.7em} +  \int_{0}^{\tau_{\mathcal{D}_{s}}} e^{-\int_{t}^{t+u} r(X_{s}) \, ds } \left( \mathcal{A}_{X} - \mathcal{A}_{\bar{X}} \right) V_{E}^{\bar{X}}(t+u,X_{u}) \, du 
\label{ESTIA}
\end{align}
is an unbiased estimator of the American-style option price $V_{A}^{X}(t,x)$. In particular, combining standard Monte Carlo techniques with (\ref{ESTIA}), provides an extension to, e.g., the ``crude'' least squares Monte Carlo technique discussed in \cite{LS01} (cf.~also \cite{Ce96}), and may considerably improve the pricing confidence when compared with this method. Discussing the circumstances under which (\ref{ESTIA}) may be highly beneficial is the content of the next section. \vspace{1em} \\
\noindent \underline{\bf Remark 2.} \vspace{0.2em} \\
Following the above derivations and combining them with the arguments provided in Remark~1, one sees that the JDOI-estimator of American-style contracts having the form of (\ref{DBgenEqua}) becomes
\begin{equation}
\mathcal{Z}_{t,x,\tau_{{D}_{s}}}^{X}  = \mathcal{Z}_{t,x,\tau_{{D}_{s}}}^{X,\ast} \mathds{1}_{ \left \{ X_{\tau_{\mathcal{D}_{s}}} \in \, \mathcal{D}_{s}^{\ast} \right\}} + \mathcal{Z}_{t,x,\tau_{{D}_{s}}}^{X,\mathcal{S}_{\partial}} \mathds{1}_{ \left \{ X_{\tau_{\mathcal{D}_{s}}} \in \, \mathcal{D}_{s}^{\mathcal{S}_{\partial}} \right\}} 
\label{ESTIADoubleTOTAL}
\end{equation}
\noindent where
\begin{align}
 \mathcal{Z}_{t,x,\tau_{{D}_{s}}}^{X,\ast} & =  V_{E}^{\bar{X}}(t,x) + e^{-\int_{t}^{t+\tau_{\mathcal{D}_{s}}} r(X_{s}) \, ds } \big(  V_{A}^{X}(t+\tau_{\mathcal{D}_{s}}, X_{\tau_{\mathcal{D}_{s}}}) - V_{E}^{\bar{X}}(t+\tau_{\mathcal{D}_{s}}, X_{\tau_{\mathcal{D}_{s}}}) \big) \hspace{9em} \nonumber \\
& \hspace{13.7em} +  \int_{0}^{\tau_{\mathcal{D}_{s}}} e^{-\int_{t}^{t+u} r(X_{s}) \, ds } \left( \mathcal{A}_{X} - \mathcal{A}_{\bar{X}} \right) V_{E}^{\bar{X}}(t+u,X_{u}) \, du ,
\label{ESTIADouble1}
\end{align}
\noindent and
\begin{equation}
\mathcal{Z}_{t,x,\tau_{{D}_{s}}}^{X,\mathcal{S}_{\partial}} =  V_{E}^{\bar{X}}(t,x) +  \int_{0}^{\tau_{\mathcal{D}_{s}}} e^{-\int_{t}^{t+u} r(X_{s}) \, ds } \left( \mathcal{A}_{X} - \mathcal{A}_{\bar{X}} \right) V_{E}^{\bar{X}}(t+u,X_{u}) \, du ,
\label{ESTIADouble2}
\end{equation}
\noindent We will make use of Equation (\ref{ESTIADoubleTOTAL}) in Section~\ref{SEC_Applications}, where we will provide an application of the JDOI method to price up-and-out put (UOP) barrier options under the the Heston$\oplus \mbox{3/2}\oplus$Jumps (H$\mbox{3/2}$J) model. \vspace{0.2em} \\
\mbox{ } \hspace{42.5em} \scalebox{0.75}{$\blacklozenge$}
\subsection{JDOI Method and Variance-Reduction}	
Having established unbiasedness of $\mathcal{Z}_{t,x,\tau_{{D}_{s}}}^{X}$, we now investigate under which conditions a low variance of this estimator is achieved. First, combining (\ref{ESTIA}) with (\ref{DERIV_Euro_Part}) and Notation (\ref{Abbrev1}) and (\ref{Abbrev2}), allows to obtain that
\begin{align}
\mathcal{Z}_{t,x,\tau_{{D}_{s}}}^{X} & := e^{-\int_{t}^{t+\tau_{\mathcal{D}_{s}}} r(X_{s}) \, ds } \, V_{A}^{X}(t+\tau_{\mathcal{D}_{s}}, X_{\tau_{\mathcal{D}_{s}}}) \hspace{10em} \nonumber \\
& \hspace{3.5em} - \sum \limits_{j=1}^{m} \int_{0}^{\tau_{\mathcal{D}_{s}}} \overline{\mathcal{R}}_{W,X}^{j}(t,u) \,dW_{u}^{j}  - \sum \limits_{k=1}^{\ell} \int_{0}^{\tau_{\mathcal{D}_{s}}} \int_{\mathbb{R}}^{} \overline{\mathcal{R}}_{N,X}^k(t,u,z_{k}) \, \tilde{N}_{k}(du,dz_{k}).
\end{align}
Additionally, the square-integrability of $(M_{u}^{A})_{u \in [t,T]}$ for any initial value $Z_{0}=(t,x)$ (cf.~\ref{Square_INT}) implies that there exist unique $\mathbf{F}$-predictable (square-integrable) representation kernels\footnote{cf.~\cite{Ku04}, \cite{Ku10} for details on (local) martingale representations.}
\begin{equation}
\big(\xi^{W}(u) \big)_{u \in [t,T]} = \big\{ \xi^{W}(u) = \big(\xi_{1}^{W}(u), \ldots, \xi_{m}^{W}(u)\big): \, u \in [t,T] \big \}, \label{REP_KERNEL1}
\end{equation}
\begin{equation}
\big(\xi^{N}(u,z) \big)_{u \in [t,T], z \in \mathbb{R}^{\ell}} = \big\{ \xi^{N}(u,z) = \big(\xi_{1}^{N}(u,z_{1}), \ldots, \xi_{\ell}^{N}(u,z_{\ell})\big): \, u \in [t,T], z \in \mathbb{R}^{\ell}\big \}, \label{REP_KERNEL2}
\end{equation}
such that
\begin{align}
e^{-\int_{t}^{t+\tau_{\mathcal{D}_{s}}} r(X_{s}) \, ds } \, V_{A}^{X}(t+\tau_{\mathcal{D}_{s}}, X_{\tau_{\mathcal{D}_{s}}}) & = V_{A}^{X}(t, x) + \sum \limits_{j=1}^{m} \int_{0}^{\tau_{\mathcal{D}_{s}}} \xi_{j}^{W}(u) \,dW_{u}^{j} + \sum \limits_{k=1}^{\ell} \int_{0}^{\tau_{\mathcal{D}_{s}}} \int_{\mathbb{R}}^{} \, \xi_{k}^{N}(u,z_{k}) \, \tilde{N}_{k}(du,dz_{k}).
\end{align}
Therefore, using the generalized Itô isometry (cf.~\cite{Ap09}), we arrive at the following result
\begin{align}
\mbox{Var} \big(\mathcal{Z}_{t,x,\tau_{{D}_{s}}}^{X} \big) & = \mathbb{E}_{t,x}^{\mathbb{Q}} \Bigg[ \bigg( \sum \limits_{j=1}^{m} \int_{0}^{\tau_{\mathcal{D}_{s}}} \big( \xi_{j}^{W}(u) - \overline{\mathcal{R}}_{W,X}^{j}(t,u) \big) \,dW_{u}^{j} \nonumber \\
& \hspace{10em} + \sum \limits_{k=1}^{\ell} \int_{0}^{\tau_{\mathcal{D}_{s}}} \int_{\mathbb{R}}^{} \, \big(\xi_{k}^{N}(u,z_{k}) - \overline{\mathcal{R}}_{N,X}^k(t,u,z_{k}) \big) \, \tilde{N}_{k}(du,dz_{k}) \bigg)^{2} \Bigg] \nonumber \\
& = \sum \limits_{j=1}^{m} \int_{0}^{T-t} \mathbb{E}_{t,x}^{\mathbb{Q}} \Big[ \mathds{1}_{\{ u < \tau_{\mathcal{D}_{s} \}}} \big( \xi_{j}^{W}(u) - \overline{\mathcal{R}}_{W,X}^{j}(t,u) \big)^{2} \Big] \, du \nonumber \\
& \hspace{10em} + \sum \limits_{k=1}^{\ell} \int_{0}^{T-t} \int_{\mathbb{R}}^{} \, \mathbb{E}_{t,x}^{\mathbb{Q}} \Big[\mathds{1}_{\{ u < \tau_{\mathcal{D}_{s}} \}}  \big(\xi_{k}^{N}(u,z_{k}) - \overline{\mathcal{R}}_{N,X}^k(t,u,z_{k}) \big)^{2} \Big] \,\Pi_{k}(z_{k}) \, du.
\end{align}
In particular, if $u \mapsto e^{-\int_{t}^{t+u} r(X_{s}) \, ds } \, V_{A}^{X}(t+u, X_{u})$ is sufficiently smooth, an application of Itô's formula gives that
\begin{equation}
\xi_{j}^{W}(u) = \sum \limits_{i=1}^{d} e^{-\int_{t}^{t+u} r(X_{s}) \, ds} \,\sigma^{ij}(X_{u}) \, \partial_{x_{i}} V_{A}^{X}(t+u,X_{u}) =: \mathcal{R}_{W,X}^{j}(t,u),
\end{equation}
\begin{equation}
\xi_{k}^{N}(u,z_{k}) = e^{-\int_{t}^{t+u} r(X_{s}) \, ds} \Big[ V_{A}^{X}\big(t+u,X_{u-} + \gamma^{\star,k}(X_{u-},z_{k}) \big) - V_{A}^{X}(t+u,X_{u-}) \Big] =: \mathcal{R}_{N,X}^k(t,u,z_{k}),
\end{equation}
and this finally shows that the variance will be especially small, whenever for all $j \in \{1, \ldots, m \}$ and $k \in \{1, \ldots, \ell \}$ the functional differences $\mathcal{R}_{W,X}^{j}(\cdot) - \overline{\mathcal{R}}_{W,X}^{j}(\cdot)$ and $\mathcal{R}_{N,X}^k(\cdot) - \overline{\mathcal{R}}_{N,X}^k(\cdot)$, are limited. This is in line with the discussion provided for pure diffusions in \cite{HP02}.

\subsection{Stopping Time Algorithms}
\label{StoppingTime_Section}
\noindent Starting with the optimal stopping policy (\ref{OptimalStopping}) and the resulting representation of the value function~(\ref{REPREimp}), we provided a detailed discussion of our JDOI method for American-style options in the last sections. However, computing the optimal stopping time \eqref{OptimalStopping} in practical application is not an evident task and several algorithms have been proposed in the literature. One standard approach is the widely used \gls{lsmc} algorithm -- originally proposed by Carriere in \cite{Ce96} and popularized by Longstaff and Schwartz in their seminal 2001 paper \cite{LS01} -- that relies on simulated sample paths of the underlying process at discrete points in time to recursively estimate the continuation value at each time step. Due to its ubiquity in the financial industry, we will base our numerical studies of Section~\ref{SEC_numerical_results} on exactly this method, but emphasize that various other approaches -- e.g.~the use of Gaussian process regression or (deep) neural networks, as proposed in \cite{BCJ19}, \cite{BCJ20}, \cite{GMZ20}, and \cite{RW20} -- could be adopted as well. For completeness, we provide a brief outline of the LSMC method in this section.

Suppose that we are given a stochastic process $(X_{t})_{t \in [0,T]}$ and a payoff function $G(\cdot)$ satisfying all of the assumptions outlined in Section~\ref{GeneralSetting} and that we want to derive the value function $V_{A}^{X}(0,\cdot)$.\footnote{For simplicity of the exposition, we take $t = 0$ but note that the more general case of $t \in [0,T]$ can be treated analogously.}~We then consider a discretization of $N+1$ steps represented by the grid $\mathcal{T} \coloneqq \left\{ t_0, t_1, \ldots, t_N\right\}$ with $0=t_0<t_1<\ldots<t_N=T$ and solve instead the approximate, discrete version to (\ref{OptimalStopping})-(\ref{REPREimp}). From the characterization of the optimal stopping time by the continuation and stopping regions in (\ref{OptimalStopping})-(\ref{REPREimp}), the optimal stopping time problem at a given point in time $t_n\in\mathcal{T}$ can be fully characterized by a comparison of the value of continuation $V_A^X(t_n,X_{t_n})$ against the immediate payoff $G(t_n,X_{t_{n}})$ from exercising the option; in particular, the optimal policy will be to exercise and receive the immediate payoff whenever the payoff is greater than or equal to the value of continuation.

The \gls{lsmc} algorithm utilizes a recursive approach to the optimal stopping problem by exploiting the trivial solution to the stopping time problem at the time of maturity $T=t_N$, where the contract is optimally exercised if and only if the payoff is positive, hence this provides an initialization of the optimal policy at the terminal time. This allows for the construction of a linear regression estimator of the value of continuation $V_A^X(t_{N-1},X_{t_{N-1}})$ from the previous time step $t_{N-1}$ using the terminal payoff as a dependent variable and regressing on a countable set of $\mathcal{F}_{t_{N-1}}$-measurable basis functions applied to realizations of the underlying process. Many choices of basis functions are possible, e.g. the Laguerre polynomials
\begin{equation}\label{eq:laguerre}
\begin{split}
L_0 \left( x \right) &= 1, \\
L_1 \left( x \right) &= -x + 1, \\
L_2 \left( x \right) &= \frac{1}{2} \left( x^2 - 4x + 2 \right), \\
&\cdots
\end{split}
\end{equation}
The basis functions can be applied to only realizations of the underlying process $X_{t_{N-1}}$ yielding a strictly positive immediate payoff, i.e. paths such that $G(t_{N-1}, X_{t_{N-1}})>0$.\footnote{In the case of a multi-dimensional stochastic process, one may include cross-product interaction terms as well, however this may not always be necessary for all combinations of terms for the chosen basis and orders to achieve the necessary accuracy in practice and can lead to numerical instability.} Using only these paths  and restriction to paths with a positive payoff does not lead to any bias as these would not be exercised under any optimal policy.

From the estimated continuation value, the immediate payoffs at time $t_{N-1}$ can be compared against the estimated continuation value, and for all cases where the payoff from exercising exceeds this estimate, the optimal stopping policy approximation for the given paths is updated to exercise at time $t_{N-1}$.

The same steps are repeated at time $t_{N-2}$: the value of continuation is estimated via regression using the time $t_{N-1}$ value resulting from the latest iteration of the optimal stopping policy and regressors constructed from basis functions of $X_{t_{N-2}}$. Immediate payoffs at time $t_{N-2}$ are again compared against the resulting estimates, and when payoffs exceed these the policy is updated to exercise for these paths. This procedure is repeated recursively until time step $t_0=0$, producing an approximation of the optimal stopping policy across the discrete points $t_n \in \mathcal{T}$ and the cash flows resulting from following this policy.

The \gls{lsmc} method in effect approximates both the continuation value at a given time step, as well as the (possibly continuous) exercise rights by only considering exercise rights on the discrete grid $\mathcal{T}$, effectively treating an American-type option as a Bermudan option with $N+1$ equidistant exercise rights. As shown in \cite{LS01}, relative errors benchmarked against solutions from \gls{fd} methods are however already miniscule with $50$ exercise rights in the Black-Scholes case with $T=1$.

The \gls{lsmc} method has found wide applicability across industry and academia, with many subsequent articles further investigating, extending and applying the method across a wide variety of problems, see e.g. \cite{CLP02}, \cite{Eg05}, \cite{EKT07}, \cite{GY04}, \cite{CGU08} and \cite{BT04}.

% doi methods applications

\section{JDOI Applications}
\label{SEC_Applications}
Having derived and discussed the JDOI method within the full class of Lévy SDEs, our next goal consists in exemplifying its usefulness in a concrete, though very general jump-diffusion model. Our model combines the multifactor extension of Grasselli's 4/2 stochastic volatility model (cf.~\cite{Gr17}) with mixed-exponentially distributed jumps (cf.~\cite{CK11}) and has the particularity to englobe several of the most important dynamics used in option pricing theory. On the pure diffusion side, the model can be viewed as superposition of two stochastic volatility dynamics, one specified by the Heston dynamics (cf.~\cite{He93}) and the other following a 3/2 model (cf.~\cite{He97}, \cite{Pl97}). Therefore, the resulting diffusion generates an implied volatility surface that can be described by factors with different characteristics. On the pure jump side, relying on mixed-exponential distributions allows to strengthen the model's (analytical) tractability while keeping its generality. Indeed, it is well-known that mixed-exponential distributions are dense -- in the sense of weak convergence -- in the class of all distributions, and that they, therefore, offer the possibility to approximate any jump distribution (cf.~\cite{BH86}, \cite{CK11}). Our final model specification is referred to as the Heston$\oplus$3/2$\oplus$Jumps (H3/2J) market model.

\subsection{The H$\mathbf{3/2}$J Market Model}
\label{subsec.: Heston + 3/2 + Jumps Model}

Following the theory developed in Section~\ref{SEC_Method}, we consider -- under $\mathbb{Q}$ -- a financial market consisting of the (deterministic) savings account, $(B_{r}(t))_{t \in [0,T]}$, given by
\begin{equation}
B_{t}(r) := X_{t}^{0} := e^{r t}, \hspace{1.5em} r \geq 0, \, t \geq 0,
\label{Bond_Dynamics}
\end{equation}
and a $3$-dimensional stochastic process $(X_t)_{t \in [0,T]}=\{(S_t, \nu_t, \eta_t) : t \in [0,T] \}$, $X_{0} := x =(s_0,\nu_0,\eta_0) $, representing asset and state price dynamics, respectively given by
\begin{align}
\label{Asset_Dynamics}
dS_t = S_{t-} \bigg( (r-\delta-\lambda\zeta) dt + c_1\sqrt{\nu_t} dW_t^1 + c_2\sqrt{\eta_t} dW_t^2 + d\bigg( \sum \limits_{i=1}^{N_{t}} (e^{Y_{i}} - 1) \bigg) \bigg) , \hspace{1.5em} s_{0} \geq 0,
\end{align}
\noindent and 
\begin{equation}
d\nu_t  = \kappa_1(\theta_1-\nu_t) dt + \sigma_1 \sqrt{\nu_t} dW_t^\nu,  \hspace{1.5em} \nu_{0} \geq  0 ,
\label{Vol_Dynamics1}
\end{equation}
\begin{equation}
d\eta_t  = \kappa_2 (\theta_2 - \eta_t) \eta_t dt + \sigma_2 \eta_t ^{3/2} dW_t^\eta, \hspace{1.5em} \eta_{0} \geq 0 ,
\label{Vol_Dynamics2}
\end{equation}
where $\delta \in \mathbb{R}$ denotes the dividend yield, $\zeta :=\mathbb{E}^{\mathbb{Q}}\left[ e^{Y_{1}}-1 \right]$ expresses the average (percentage) jump size, $c_{1}, c_{2} \in \mathbb{R}$ and $\kappa_{1}, \kappa_{2}, \theta_{1}, \theta_{2}, \sigma_{1}, \sigma_{2} > 0$. Here, we assume that appropriate Feller conditions hold, i.e.~we require for Dynamics~(\ref{Vol_Dynamics1}) that $2 \kappa_{1} \theta_{1} \geq \sigma_{1}^{2}$ and for Dynamics~(\ref{Vol_Dynamics2}) that $2\kappa_{2}^{\star} \theta_{2}^{\star} \geq (\sigma_{2}^{\star})^{2}$ is satisfied, with $\kappa_{2}^{\star} := \kappa_{2} \theta_{2}$, $\theta_{2}^{\star} := (\kappa_{2} + \sigma_{2}^{2}) / \kappa_{2}^{\star}$, and $\sigma_{2}^{\star} := -\sigma_{2}$. Additionally, the Brownian motions $(W_{t}^{1})_{t \in [0,T]}$, $(W_{t}^{2})_{t \in [0,T]}$, $(W_{t}^{\nu})_{t \in [0,T]}$, and $(W_{t}^{\eta})_{t \in [0,T]}$ are assumed to satisfy
\begin{equation}
[W^{1}, W^{2} ]_{t} = [W^{\nu}, W^{\eta}]_{t} = 0, \hspace{1.5em} \mbox{and} \hspace{2em}  [W^{1}, W^{\nu}]_{t} = \rho_{1} t, \hspace{1em} [W^{2}, W^{\eta}]_{t} = \rho_{2} t, \hspace{1.5em} t \in [0,T],
\end{equation}
with correlation coefficients $\rho_{1}, \rho_{2} \in [-1,1]$ and $\lambda > 0$ denotes the intensity of the (independent) Poisson process $(N_{t})_{t \in [0,T]}$. Finally, the jumps $(Y_{i})_{i \in \mathbb{N}}$ are assumed to be independent of $(N_{t})_{t \geq 0}$ and to form a sequence of independent and identically distributed random variables following a mixed-exponential distribution, i.e.~their (common) density function $\varphi_{Y_{1}}^{\text{mix}}(\cdot)$ is given by
\begin{align}
\label{subsec.2.3.1:mixed-exponetial distribution function}
\varphi_{Y_{1}}^{\text{mix}}(y):= p_u\sum_{i=1}^m p_i a_i e^{-a_i y} \mathds{1}_{\{y\geq 0\}} + q_d\sum_{j=1}^n q_j b_j e^{b_j y} \mathds{1}_{\{y< 0 \}},
\end{align}
\noindent where $p_u \geq 0$, $q_d=1-p_u \geq 0$, and the remaining parameters satisfy the following conditions:
\begin{align*}
p_{i} \in \mathbb{R},   \hspace{0.5em} & \forall  i=1,\ldots,m,  \hspace{1.5em} \mbox{with} \hspace{1.5em} \sum_{i=1}^m p_i=1, \\
q_{j} \in \mathbb{R},   \hspace{0.5em} &\forall  j=1,\ldots,n,  \hspace{1.5em} \mbox{with} \hspace{1.5em} \sum_{j=1}^n q_j=1, \\
& a_{i}  > 1 ,   \hspace{0.5em}  \forall  i=1,\ldots,m, \\
& b_{j}  > 0 ,   \hspace{0.5em}  \forall  j=1,\ldots,n.
\end{align*}
Because $p_i$, $i=1, \ldots, m$, and $q_j$, $j =1, \ldots, n$, can be negative, these parameters have to satisfy certain conditions to guarantee that the function $\varphi_Y^{\text{mix}}(\cdot)$ is always nonnegative and is a probability density function. As noted in \cite{CK11}, a necessary condition for $\varphi_Y^{\text{mix}}(\cdot)$ to be a probability density function is $p_1>0$, $q_1>0$, $\sum_{i=1}^{m}p_ia_i \geq 0$, and $\sum_{j=1}^n q_j b_j \geq 0$, while a simple sufficient condition is given by $\sum_{i=1}^k p_ia_i \geq 0$ for all $k=1,\dotsc,m$ and $\sum_{j=1}^l q_j b_j \geq 0$ for all $j=1,\dotsc,n$. Lastly, we note that the condition $a_i >1$, for $i=1,\dotsc,m$, is only imposed to ensure that the stock price process $(S_t)_{t \in [0,T]}$ has finite expectation.

\subsection{An Approximate H$\mathbf{3/2}$J Market Model}
\label{H3/2Approx}
As next step towards a derivation of JDOI estimators for American standard and barrier options under the H3/2J market model, we fix and briefly discuss an approximate market model. From a practical point of view, choosing an approximate market model whose infinitesimal generator is available in analytical form is particularly desirable since this increases the computational performance -- in particular the speed -- of the resulting JDOI algorithm. Consequently, we aim to transform our H3/2J market into an approximate one with analytical greeks while still capturing the main characteristics of the H3/2J dynamics. Here, we rely on a generalized version of the (standard) Black-Scholes dynamics (GBS) that is given -- under $\mathbb{Q}$ --  by the (deterministic) savings account, as specified in (\ref{Bond_Dynamics}), and the $3$-dimensional approximation $(\bar{X}_t)_{t \in [0,T]}=\{(\bar{S}_t, \bar{\nu}_t, \bar{\eta}_t) : t \in [0,T] \}$, $\bar{X}_{0} := x =(s_0,\nu_0,\eta_0) $, having the form
\begin{equation}
\label{Asset_DynamicsGBS}
d\bar{S}_t = \bar{S}_{t} \bigg( (r-\delta) dt + \sqrt{\big(c_1^{2}\bar{\nu}_t + c_2^{2}\bar{\eta}_t}\big) dW_t \bigg) , \hspace{1.5em} s_{0} \geq 0,
\end{equation} 
\begin{equation}
d\bar{\nu}_t  = \kappa_1(\theta_1-\bar{\nu}_t) dt ,  \hspace{1.5em} \nu_{0} \geq  0 ,
\label{Vol_DynamicsGBS1}
\end{equation}
\begin{equation}
d\bar{\eta}_t  = \kappa_2 (\theta_2 - \bar{\eta}_t) \bar{\eta}_t dt, \hspace{1.5em} \eta_{0} \geq 0 ,
\label{Vol_DynamicsGBS2}
\end{equation}
where $(W_{t})_{t \in [0,T]}$ denotes another Brownian motion. Therefore, under the above GBS market version, the stochastic volatility dynamics of the H3/2J market model are approximated by their deterministic trends and the process' infinitesimal generator is obtained,  for any sufficiently well-behaved function $V:[0,T] \times \mathbb{R}^{3} \rightarrow \mathbb{R}$, as 
\begin{align}
\mathcal{A}_{\bar{X}} V(t,s_{0},\nu_{0}, \eta_{0}) & =(r-\delta) s_{0} \partial_{S}V(t,s_0,\nu_0,\eta_0) + \frac{1}{2} \big(c_1^2 \nu_{0} + c_2^2 \eta_{0} \big) s_{0}^2 \partial_{S}^{2} V(t,s_0,\nu_0,\eta_0)  \nonumber \\
& \hspace{4.5em}  +\kappa_1(\theta_1-\nu_{0}) \partial_{\nu}V(t,s_0,\nu_0,\eta_0) + \kappa_2(\theta_2-\eta_{0})\eta_{0}  \partial_{\eta} V(t,s_0,\nu_0,\eta_0).
\end{align}
Additionally, ODEs (\ref{Vol_DynamicsGBS1}) and (\ref{Vol_DynamicsGBS2}) can be solved to obtain that
\begin{equation}
\bar{\nu}_{t} = \theta_{1} + ( \nu_{0} - \theta_{1}) e^{-\kappa_{1} t},
\end{equation}
\begin{equation}
\bar{\eta}_{t} = \bigg( \frac{1}{\theta_{2}} + \bigg( \frac{1}{\eta_{0}} - \frac{1}{\theta_{2}}\bigg) e^{-\kappa_{2} \theta_{2} t }\bigg)^{-1},
\end{equation}
and upon combining the Dambis-Dubins-Schwarz theorem (c.f~\cite{RY99}) with the deterministic variance, $(\bar{\sigma}_{t}^{2})_{t \in [0,T]}$, given, for $t \in [0,T]$, by
\begin{align}
\bar{\sigma}_{t}^{2} := \frac{1}{t} \int_{0}^{t} \big( c_{1}^{2} \bar{\nu}_{s} + c_{2}^{2} \bar{\eta}_{s} \big) \, ds = c_{1}^{2} \theta_{1} + c_{2}^{2} \theta_{2} + \frac{c_{1}^{2}}{\kappa_{1} t} ( \nu_{0} - \theta_{1}) (1-e^{-\kappa_{1}t}) + \frac{c_{2}^{2}}{\kappa_{2} t } \log \bigg( \frac{\eta_{0}}{\theta_{2}} + \bigg( 1 - \frac{\eta_{0}}{\theta_{2}}\bigg) e^{-\kappa_{2} \theta_{2} t} \bigg),
\label{Deter_Variance}
\end{align}
we can rewrite Dynamics (\ref{Asset_DynamicsGBS})-(\ref{Vol_DynamicsGBS2}) in the standard Black-Scholes form
\begin{equation}
d\bar{S}_t = \bar{S}_{t} \bigg( (r-\delta) dt + \bar{\sigma}_{t} dW_t \bigg) , \hspace{1.5em} s_{0} \geq 0.
\label{Asset_Dynamics_Rewritten}
\end{equation}
Consequently, numerous expressions are available in this simplified market environment in analytical form and this substantially reduces the computational costs associated with our final JDOI estimates (cf.~Sections~\ref{StandardAmerOptions} and \ref{AmerBarriers}).

\subsection{Standard American Options under the H$\mathbf{3/2}$J Model}
\label{StandardAmerOptions}
We now turn to concrete applications of the JDOI method under the H3/2J market model and start by dealing with the case of standard American options. Here, we first recall that the price of a European put option written on the approximate (GBS-)dynamics (\ref{Asset_DynamicsGBS})-(\ref{Vol_DynamicsGBS2}) is directly given in terms of the standard Black~\&~Scholes formula (cf.~(\ref{Asset_Dynamics_Rewritten})) and reads
\begin{equation}
V_{E}^{\bar{X},\mathcal{P}}(t,s_{0},\nu_{0},\eta_{0};K)  = Ke^{-r(T-t)} \mathcal{N} \bigg( -d_{2} \bigg(\frac{s_{0}}{K}, \bar{\sigma}_{T-t}, T-t \bigg) \bigg) - s_{0}e^{-\delta (T-t)} \mathcal{N} \bigg( -d_{1} \bigg(\frac{s_{0}}{K}, \bar{\sigma}_{T-t}, T-t \bigg)\bigg),
\label{Put_BlackScholes}
\end{equation}  
\noindent where $\mathcal{N}(\cdot)$ denotes, as usual, the standard normal cumulative distribution function, and
\begin{equation}
\label{Def_d1d2}
d_{1}(\chi,\zeta,\xi) := \frac{\log(\chi) + \left(r-\delta + \frac{1}{2}\zeta^{2}\right) \xi}{\zeta\sqrt{\xi}}, \hspace{2.5em} d_{2}(\chi, \zeta, \xi) := d_{1}(\chi,\zeta,\xi) - \zeta \sqrt{\xi}.
\end{equation}
\noindent Furthermore, the operator difference in (\ref{ESTIA}) takes the form
\begin{align}
\label{Difference_InfGen}
\left(\mathcal{A}_{X} - \mathcal{A}_{\bar{X}}\right)& V_{E}^{\bar{X},\mathcal{P}}(t,s_0,\nu_0,\eta_0;K) \nonumber \\
\nonumber =  & -\lambda \zeta s_{0}  \partial_S V_{E}^{\bar{X},\mathcal{P}}(t,s_0,\nu_0,\eta_0;K) + \frac{1}{2}\sigma_1^2\nu_{0} \partial_{\nu}^{2}V_{E}^{\bar{X},\mathcal{P}}(t,s_0,\nu_0,\eta_0;K)\\
 &+ \rho_1 s_{0} c_1\sigma_1\nu_{0}\partial_{S}\partial_{\nu}V_{E}^{\bar{X},\mathcal{P}}(t,s_0,\nu_0,\eta_0;K)\\
\nonumber&+ \frac{1}{2}\sigma_2^2\eta_{0}^3\partial_{\eta}^{2}V_{E}^{\bar{X},\mathcal{P}}(t,s_0,\nu_0,\eta_0;K) +\rho_2s_{0}c_2\sigma_2\eta_{0}^{2}\partial_{S}\partial_{\eta}V_{E}^{\bar{X},\mathcal{P}}(t,s_0,\nu_0,\eta_0;K)\\
\nonumber &+ \lambda \int_\mathbb{R}V_{E}^{\bar{X},\mathcal{P}}(t,s_0e^y,\nu_0,\eta_0;K)\varphi_Y^{\text{mix}}(y)\mathrm{d}y
- \lambda V_{E}^{\bar{X},\mathcal{P}}(t,s_0,\nu_0,\eta_0;K),
\end{align}
and the main difficulty in the computation of the JDOI estimator consists in obtaining an analytical expression for the integral term in (\ref{Difference_InfGen}). Indeed, due to the form of the European put in (\ref{Put_BlackScholes}), standard Black~\&~Scholes Greeks can be combined with the derivatives of the (deterministic) variance process (\ref{Deter_Variance}) to arrive at the remaining Greeks in (\ref{Difference_InfGen}).\footnote{For the sake of completeness, these expressions are provided as part of~Appendix~A.}~For the derivation of the integral, we combine the advantage of mixture distributions with the nice properties of simple exponential densities to arrive at the following result: 
\begin{align}
 \int_\mathbb{R} V_{E}^{\bar{X},\mathcal{P}}(t,s_0e^y,\nu_0,\eta_0;K)\varphi_Y^{\text{mix}}(y)dy  \hspace{20em} \nonumber \\
  \hspace{5em} = Ke^{-r(T-t)} \Psi_2\bigg(\frac{s_0}{K},\bar{\sigma}_{T-t},T-t\bigg)- s_0e^{-\delta (T-t)}\Psi_1 \bigg(\frac{s_0}{K},\bar{\sigma}_{T-t},T-t\bigg),
\label{Integral_StandardOption}
\end{align}
\noindent Here, the functions $\Psi_{1}(\cdot)$ and $\Psi_{2}(\cdot)$ are defined as in (\ref{PSI_1_first}) and (\ref{PSI_1_second}) and the (full) derivation of these results is provided in Appendix~A. Consequently, fixing a stopping time algorithm\footnote{As discussed in Section~\ref{StoppingTime_Section}, we will rely on the Longstaff-Schwarz method (cf.~\cite{Ce96}, \cite{LS01}).}~and combining it with all these analytical expressions leaves us with a straightforward and efficient implementation of the JDOI method for standard American options (cf.~(\ref{ESTIA})).

\subsection{American Barrier Options under the H$\mathbf{3/2}$J Model}
\label{AmerBarriers}
\noindent As a last step, we apply our JDOI extension to price American barrier options under the H3/2J market model. Here, we focus on (regular) up-and-out put (UOP) barrier options but emphasize that the exact same techniques could be applied to any other type of barrier contracts. \vspace{1em} \\
\noindent To start, we note that closed-form solutions for European barrier options have been derived under the classical Black-Scholes dynamics (cf.~among others~\cite{RR91}, \cite{JY06}) and that these results can be naturally extended to our GBS framework. For instance, the value of a (regular) European up-and-out put option under the GBS model is given, with $ x = (s_{0}, \nu_{0}, \eta_{0}) \in \left( \mathbb{R}_{0}^{+} \right)^{3}$, by	
\begin{align}
V_{E}^{\bar{X},\mathcal{UOP}}(t,s_{0},\nu_{0},\eta_{0}; K,H) & = \mathbb{E}_{t,x}^{\mathbb{Q}} \left[ e^{-r(T-t)} \, G(T,\bar{X}_{T-t}) \right] \nonumber \\
& = \mathbb{E}_{t,x}^{\mathbb{Q}} \left[ e^{-r (T-t)} \, (K - \bar{S}_{T-t})^{+} \, \mathds{1}_{\{ \tau_{H} > T-t \}} \right] \nonumber \\
& = V_{E}^{\bar{X},\mathcal{P}} (t,s_{0},\nu_{0},\eta_{0};K) - \left( \frac{H}{s_{0}}\right)^{2(\gamma - 1)} V_{E}^{\bar{X},\mathcal{P}}\bigg(t,\frac{H^2}{s_{0}},\nu_{0},\eta_{0};K \bigg).
\label{Def_UOP}
\end{align} 
\noindent Here, $\tau_{H} := \inf \{ t \geq 0: \, \bar{S}_{t} \geq H \}$ denotes the first-passage time of the underlying asset $(S_{t})_{t \geq 0}$ above the (upper) barrier level $H$ and we have additionally set $\gamma := \frac{r-\delta}{\bar{\sigma}_{T-t}^{2}} + \frac{1}{2}$. Combining this decomposition in terms of standard GBS put options with the derivations of Section~\ref{StandardAmerOptions} (cf.~Appendix~A) directly allows for a recovery of all the derivative expressions in (\ref{Difference_InfGen}). These expressions are provided as part of Appendix~B and we are consequently left again with a derivation of the integral term in (\ref{Difference_InfGen}).  For this integral term, we see that
\begin{align}
\int_{\mathbb{R}} & V_{E}^{\bar{X},\mathcal{UOP}}(t,s_{0}e^{y},\nu_{0},\eta_{0}; K,H) \, \varphi_{Y}^{mix}(y) \, dy  \nonumber \\
& \hspace{0.6em} = \underbrace{\int_{\mathbb{R}} V_{E}^{\bar{X},\mathcal{P}}(t,s_{0}e^{y},\nu_{0},\eta_{0}; K) \, \varphi_{Y}^{mix}(y) \, dy}_{\text{(I)}}  - \left( \frac{H}{s_{0}}\right)^{2(\gamma - 1)} \underbrace{\int_{\mathbb{R}} V_{E}^{\bar{X},\mathcal{P}}\bigg(t,\frac{H^2}{s_{0}e^{y}},\nu_{0},\eta_{0};K \bigg) \,e^{-2(\gamma - 1)y} \, \varphi_{Y}^{mix}(y) \, dy}_{\text{(II)}} \label{UOP_Int_TERM}
\end{align}
\noindent and note that we already dealt with Integral (I), as part of Section~\ref{StandardAmerOptions}, in~(\ref{Integral_StandardOption}). Therefore, we are left with the derivation of Integral (II) and following the same techniques as the ones used in Appendix~A while imposing slightly stronger integrability conditions (cf.~Equation~(\ref{Int_COND_STRONG})), finally gives the following result:
\begin{align}
\int_{\mathbb{R}} & V_{E}^{\bar{X},\mathcal{P}}\bigg(t,\frac{H^2}{s_{0}e^{y}},\nu_{0},\eta_{0};K \bigg) \,e^{-2(\gamma - 1)y} \, \varphi_{Y}^{mix}(y) \, dy \nonumber \\
& \hspace{2.5em} = K e^{-r(T-t)} \Psi_{B,2}\bigg(\frac{H^{2}}{s_{0} K }, \bar{\sigma}_{T-t}, T-t , \gamma \bigg) - \frac{H^2}{s_{0}} e^{-\delta (T-t)} \Psi_{B,1}\bigg(\frac{H^{2}}{s_{0} K }, \bar{\sigma}_{T-t}, T-t , \gamma \bigg).
\end{align}
\noindent Here, the functions $\Psi_{B,1}(\cdot)$ and $\Psi_{B,2}(\cdot)$ are given as in (\ref{PSI_2_first}) and (\ref{PSI_2_second}) and the derivation of this expression is provided in Appendix~B. Together with Representation~(\ref{Integral_StandardOption}), this provides us with an analytical expression for (\ref{UOP_Int_TERM}) and combining all these results leaves us with a straightforward and efficient implementation of the JDOI method for American barrier options (cf.~(\ref{ESTIA})).

% numerical results

\section{Numerical Results}
\label{SEC_numerical_results}
Based on the derivations in the previous sections, we now present numerical results for the implementation of the estimator in \eqref{ESTIADoubleTOTAL} for the European and American-style standard put options and up-and-put put barrier options.

\noindent For the model parameters associated with the dynamics in Section~\ref{subsec.: Heston + 3/2 + Jumps Model}, we use the values noted in Table~\ref{table:parameters} throughout all runs of the estimator unless otherwise specified. In particular, we rely on double-exponentially distributed jumps, i.e.~we choose a jump distribution \eqref{subsec.2.3.1:mixed-exponetial distribution function} with one positive and one negative jump, i.e. $n=m=1$, $p\coloneqq p_up_1$ and $q\coloneqq q_dq_1$. We then have from the moment-generating function for the exponential distribution and the definition of $\zeta$ in Section~\ref{subsec.: Heston + 3/2 + Jumps Model} that
\begin{equation}
\zeta = p\frac{a}{a-1} + q\frac{b}{b+1} - 1.
\end{equation}

\begin{table*}[h]\centering
\begin{tabular}{cccccccccccccccccc}
\toprule
$S_0$ & $\nu_0$ & $\eta_0$ & $r$  & $d$  & $\kappa_1$ & $\kappa_2$ & $\theta_1$ & $\theta_2$ & $\sigma_1$ & $\sigma_2$ & $\rho_1$ & $\rho_2$ & $a$ & $b$ & $p$ & $\lambda$ & $T$ \\ \midrule
100   & 0.01    & 0.01     & 0.04 & 0    & 0.6        & 60         & 0.01       & 0.01       & 0.1        & 10         & -0.15    & 0.15     & 100 & 25  & 0.3 & 5         & 0.5 \\
\bottomrule
\end{tabular}
\caption{Parameter set used across estimations.\label{table:parameters}}
\end{table*}

The choice of parameters are based on the Heath-Platen paper \cite{HP02} scaled down by $(0.5)^{2}=0.25$ such that the parameters for the $\nu$ and $\eta$ processes lead to a match of the total long-term mean volatility in the original paper. Additionally, $\kappa_2=\frac{\kappa_1}{\theta_2}$ is chosen to match the mean reversion rate of the $\nu$ process when starting from the long-term mean, while $\sigma_2=\frac{\sigma_1}{\eta_0}$ is chosen such that the initial volatilities of the $\eta$ and $\nu$ processes are identical. The parametrization of the jumps is inspired by the (various) configurations used in the existing literature on double-exponential jumps (cf.~\cite{CK11}, \cite{CK12}, \cite{LV17}, \cite{CV18}, \cite{FM20}).

For the \gls{lsmc} approximation of the optimal stopping time policy, we regress on Laguerre polynomials as in \eqref{eq:laguerre} up to the second order for the underlying and linear terms for the variance processes. Cross-terms are not utilized due to numerical stability considerations and the limited impact on the optimal stopping decision from interactions of the variance processes.

\begin{figure}[H]\centering
\includegraphics[width=.53\linewidth, trim=120 250 120 240, clip]{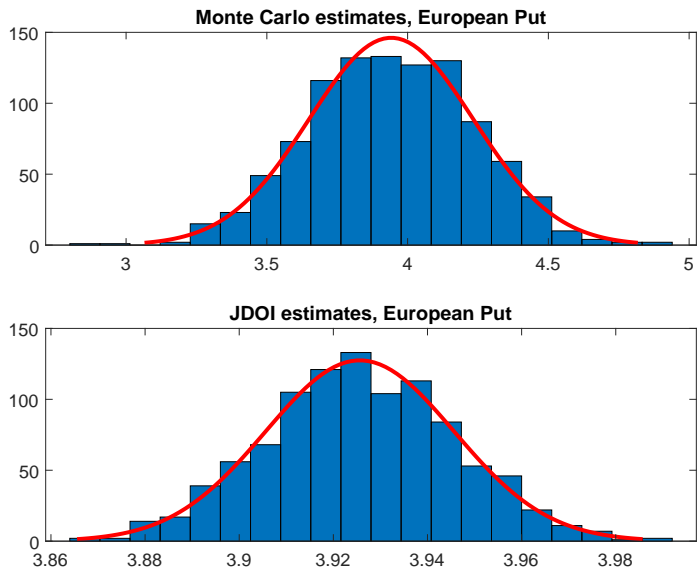}
\caption{Histogram of $1,000$ MC and JDOI estimates with $100$ time steps and $200$ sample paths for the European put option.\label{fig:euro_put_histogram}}
\end{figure}

Histograms of price estimates for the European put option for the MC and JDOI estimators respectively are presented in Figure~\ref{fig:euro_put_histogram} with strike $K=100$. The estimators are run $200$ times each, using $200$ sample paths per run and $100$ time steps per path. This provides a clear visual representation of the dramatic variance reduction achieved with the JDOI estimator; for the same number of sample paths, the JDOI estimates are closely clustered around the mean in absolute terms compared to the MC estimates.

While both the MC and JDOI estimates have means of around $3.93$, the MC estimates standard deviation is $0.3$ compared to only $0.02$ of the JDOI estimates. In other words, we can estimate improvements in expected relative errors beyond an order of magnitude for the same number of sample paths for the JDOI estimator compared to the MC estimator.

\begin{figure}[H]\centering
\includegraphics[width=.53\linewidth, trim=120 250 120 240, clip]{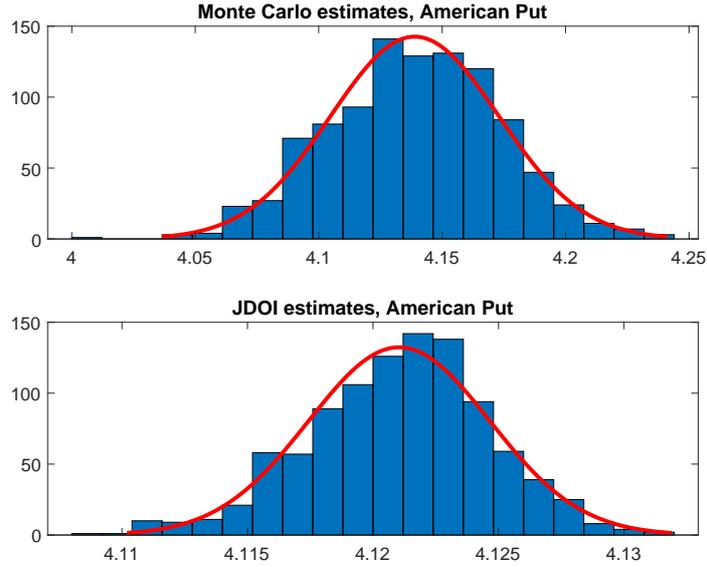}
\caption{Histogram of $1,000$ MC and JDOI estimates with $100$ time steps and $10,000$ sample paths for the American put option.\label{fig:amer_put_histogram}}
\end{figure}

An analogous histogram is presented for the American put option from Section~\ref{StandardAmerOptions} in Figure~\ref{fig:amer_put_histogram} using the same number of time steps and number of runs, but with $10,000$ sample paths per run. This is due to the added complexity in having to estimate the optimal stopping time solution based on the realized payoffs, which leads to a bottleneck in lower-sample estimation and may introduce a bias stemming from following a suboptimal exercise policy for very low sample paths.

Despite the reliance on the realized payoffs with no variance reduction in the \gls{lsmc} optimal exercise approximation, the JDOI estimates are again clustered much closer around their mean compared to the standard MC estimates. In particular, the MC mean estimates are around $4.13$ with a standard deviation of $0.034$, while the JDOI mean estimates are $4.12$ with a standard deviation of only $0.004$. While the degree of variance reduction remains impressive, it is noticeably smaller than the European case due to the aforementioned optimal stopping approximation, as well as the reliance on the European value function as an approximating payoff, since the diffusion operator cannot be derived analytically in the American case, even for the Black-Scholes case.

\begin{figure}[H]\centering
\includegraphics[width=.53\linewidth, trim=120 250 120 240, clip]{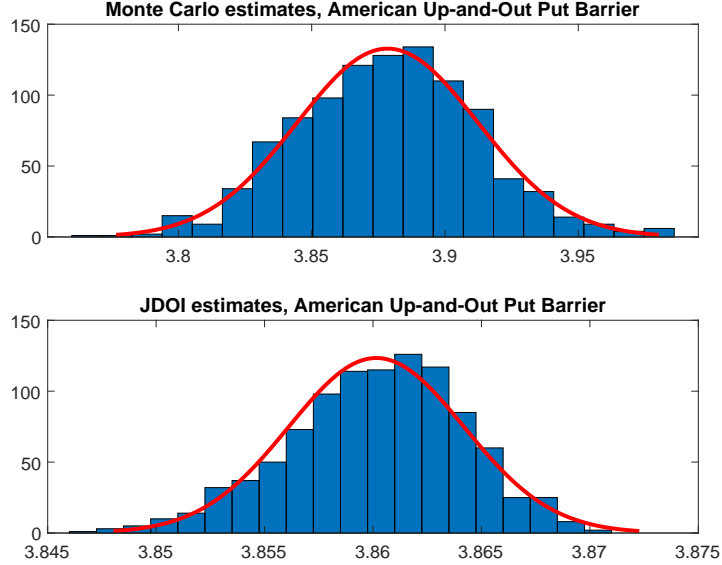}
\caption{Histogram of $1,000$ MC and JDOI estimates with $100$ time steps and $10,000$ sample paths for the American up-and-out put barrier.\label{fig:amer_uop_histogram}}
\end{figure}

We finally repeat the exercise for the main pricing problem of the article, namely the pricing of the American up-and-out put barrier from Section~\ref{AmerBarriers}. We again use a strike of $K=100$ and a barrier of $H=110$, and the same combination of sample paths and time steps as for the American put option before. The resulting histogram in Figure~\ref{fig:amer_uop_histogram} showcases comparable variance reduction as for the standard American put.

In particular, the MC estimates have a mean of around $3.87$ and standard deviation of $0.034$, while the JDOI estimates have a mean of $3.86$ and standard deviation of $0.004$, which is closely aligned with the American put case from before. The analysis of the estimator for the American up-and-out put barrier is extended further in the remainder of this section.

\begin{figure}[H]\centering
\includegraphics[width=.53\linewidth, trim=40 250 40 250, clip]{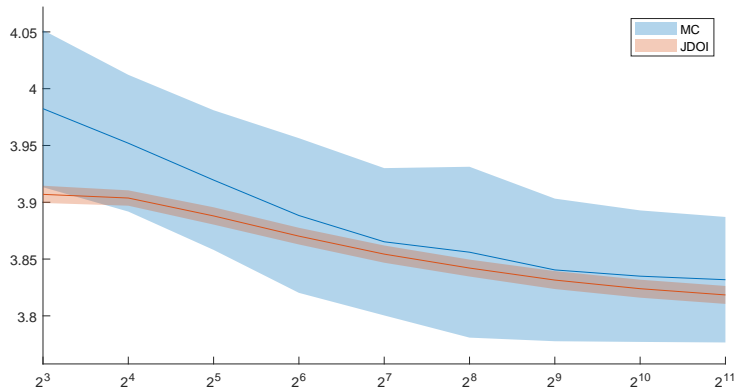}
\caption{MC and JDOI mean estimates and 95\% confidence bands from $100$ runs with varying time steps and $10,000$ sample paths for the American up-and-out put barrier.\label{fig:amer_uop_time_steps}}
\end{figure}

Next we assess the scaling of the estimator performance in number of time steps of the discretization and sample paths. We here consider only the American up-and-out barrier contract, again with strike $K=100$ and barrier $H=110$. We start by assessing the magnitude of bias from the discretization by varying the number of steps with a fixed number of sample paths at $10,000$. The resulting mean estimates and 95\% confidence bands from $100$ runs of each estimator with time steps $2^n$ for $n=3,4,\ldots,11$ are presented in Figure~\ref{fig:amer_uop_time_steps}.

From Figure~\ref{fig:amer_uop_time_steps} we can observe that the mean estimates are slightly decreasing in the granularity of the discretization for both estimators. The discretization in terms of exercise rights would imply a negative bias for a coarse grid, since exercise rights on a countable, finite grid are a special case of continuous exercise rights. On the other hand, there is a monitoring bias from only observing the underlying process at discrete times, leading to a positive bias for coarser grids in this case, as any potential knock-out from the process reaching the barrier cannot be observed between two discrete points.

In this case, one may conjecture that the positive bias from not observing knock-outs dominates the negative bias from restricting the continuous exercise rights, however the additional dependency on a stopping rule approximation, which in turn depends on the number of sample paths, complicates such direct conclusions from this observation. While methods exist to correct for the lack of observing knockouts, we do not treat this monitoring bias problem for barrier options in greater detail and instead refer to \cite{MA02} and \cite{Go09} for a further discussion.

\begin{figure}[H]\centering
\includegraphics[width=.53\linewidth, trim=40 250 40 250, clip]{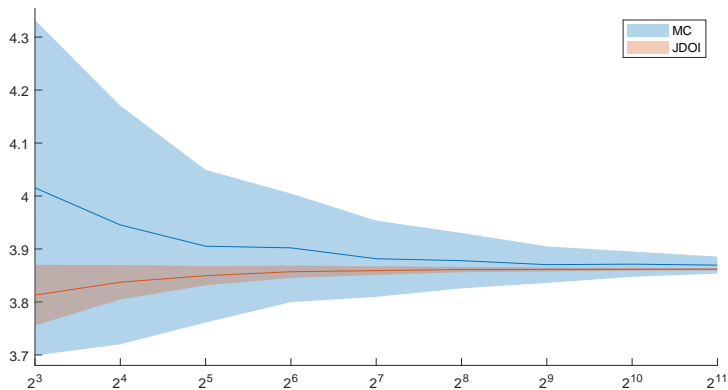}
\caption{MC and JDOI mean estimates and 95\% confidence bands from $100$ runs with varying sample paths and $100$ time steps for the American up-and-out put barrier.\label{fig:amer_uop_sample_paths}}
\end{figure}

Figure~\ref{fig:amer_uop_sample_paths} presents mean estimates and 95\% confidence bands for a fixed number of $100$ time steps over a varying number of sample paths $2^n$ for $n=9,10,\ldots,17$. In contrast to the time step analysis, the mean estimates for the JDOI estimator are now increasing in the number of sample paths, while the standard MC mean estimates are decreasing. Such biases may result from the previously discussed optimal stopping time approximation, however we here note that even for the smaller number of sample paths, the JDOI estimates are numerically close to the large-sample mean of around $3.86$ for both estimators.

Finally we present JDOI price estimates across a range of model parameters in Table~\ref{table:price_estimates} for the American up-and-out put barrier, in particular varying initial values $S_0$ of the underlying and barriers $H$, and otherwise using the parameters from Table~\ref{table:parameters}. Each price estimate is based on $100$ runs of the estimators with $10,000$ sample paths and $100$ time steps and includes confidence intervals and minimum / maximum observed values. The JDOI method consistently has extremal estimates closely aligned with the mean estimate in contrast to the MC estimator, despite following the same stopping policy for each sample, and the improvement in standard deviation is largely aligned with the previous results.

\begin{table*}[h]\centering
\begin{tabular}{llrrcrrrrcrr}
\toprule
& & \multicolumn{5}{c}{MC} & \multicolumn{5}{c}{JDOI} \\\cmidrule(lr){3-7}\cmidrule(lr){8-12}
$S_0$ & $H$ & Mean & StdDev & ConfInterval & Min & Max & Mean & StdDev & ConfInterval & Min & Max \\\midrule
90 & 110 & 10.263 & 0.0340 & [10.20,10.33] & 10.18 & 10.34 & 10.232 & 0.0055 & [10.22,10.24] & 10.22 & 10.25 \\
90 & 115 & 10.272 & 0.0346 & [10.20,10.34] & 10.16 & 10.35 & 10.251 & 0.0062 & [10.24,10.26] & 10.23 & 10.26 \\
90 & 120 & 10.279 & 0.0341 & [10.21,10.35] & 10.16 & 10.35 & 10.253 & 0.0049 & [10.24,10.26] & 10.24 & 10.26 \vspace{0.5em}\\
95 & 110 & 6.583 & 0.0325 & [6.52,6.65] & 6.50 & 6.67 & 6.565 & 0.0050 & [6.56,6.57] & 6.55 & 6.58 \\
95 & 115 & 6.657 & 0.0360 & [6.59,6.73] & 6.58 & 6.73 & 6.640 & 0.0051 & [6.63,6.65] & 6.62 & 6.65 \\
95 & 120 & 6.665 & 0.0364 & [6.59,6.74] & 6.58 & 6.78 & 6.651 & 0.0051 & [6.64,6.66] & 6.64 & 6.66 \vspace{0.5em}\\
100 & 110 & 3.878 & 0.0292 & [3.82,3.94] & 3.82 & 3.96 & 3.861 & 0.0043 & [3.85,3.87] & 3.85 & 3.87 \\
100 & 115 & 4.090 & 0.0308 & [4.03,4.15] & 4.02 & 4.18 & 4.072 & 0.0032 & [4.07,4.08] & 4.06 & 4.08 \\
100 & 120 & 4.129 & 0.0397 & [4.05,4.21] & 4.02 & 4.22 & 4.113 & 0.0031 & [4.11,4.12] & 4.11 & 4.12 \vspace{0.5em}\\
105 & 110 & 1.857 & 0.0282 & [1.80,1.91] & 1.77 & 1.92 & 1.840 & 0.0032 & [1.83,1.85] & 1.83 & 1.85 \\
105 & 115 & 2.332 & 0.0284 & [2.28,2.39] & 2.27 & 2.39 & 2.315 & 0.0031 & [2.31,2.32] & 2.31 & 2.32 \\
105 & 120 & 2.446 & 0.0303 & [2.39,2.51] & 2.36 & 2.50 & 2.429 & 0.0029 & [2.42,2.43] & 2.42 & 2.43 \vspace{0.5em}\\
110 & 110 & 0.000 & 0.0000 & [0.00,0.00] & 0.00 & 0.00 & 0.000 & 0.0000 & [0.00,0.00] & 0.00 & 0.00 \\
110 & 115 & 1.106 & 0.0212 & [1.06,1.15] & 1.05 & 1.16 & 1.088 & 0.0029 & [1.08,1.09] & 1.08 & 1.09 \\
110 & 120 & 1.368 & 0.0226 & [1.32,1.41] & 1.32 & 1.43 & 1.347 & 0.0030 & [1.34,1.35] & 1.34 & 1.35 \\
\bottomrule
\end{tabular}
\caption{Table of MC and JDOI estimates across starting values $S_0$ and barriers $H$ for the American up-and-out put barrier with mean estimates, standard deviation, $95\%$ confidence intervals, minimum and maximum observed estimates. The estimators were run $100$ times each per parameter combination with $100$ time steps.\label{table:price_estimates}}
\end{table*}

Based on the multiple benchmarks performed, we confirm that the strong variance reduction of the diffusion operator approach holds in our extension to an extended Heston model with multiple stochastic volatility drivers and double-exponential jumps. We additionaly demonstrated the succesful combination of the estimator with the \gls{lsmc} method for pricing American-style options.

While the variance reduction remains impressive in the treatment of American options, the magnitude of variance reduction is less than the European case, while a higher number of sample paths are needed to reach an adequate approximation of the optimal stopping policy. This demonstrates the value of potential future research into more efficient stopping time algorithms, as this could lead to even stronger variance reduction for the JDOI estimator, in particular for the case with few sample paths where \gls{lsmc} would no longer be feasible.

% conclusion

\section{Conclusion}
\label{SEC_Conclu}
\noindent The present article extended the literature on Monte Carlo based option pricing in several directions. On the theoretical side, we revisited the Diffusion Operator Integral (DOI) variance reduction technique originally proposed in \cite{HP02} and extended its concepts to the pricing of American-style options under (time-homogeneous) Lévy stochastic differential equations (SDEs). The resulting JDOI method can help speeding up Monte Carlo based (pricing) algorithms and allows for a great flexility in the choice of the underlying dynamics. In particular, while the existing DOI literature (cf.~\cite{HP02}, \cite{HP14}, \cite{CK18}, \cite{CKD19}) only focuses on (certain) pure diffusion dynamics, our Lévy framework has the advantage to englobe a wider range of the most frequent choices of financial models. On the application side, we provided an extensive theoretical treatment of the JDOI method for American vanilla and barrier options under the H3/2J market dynamics, a broad class of market models that combine the multifactor extension of Grasselli's 4/2 stochastic volatility model (cf.~\cite{Gr17}) with mixed-exponentially distributed jumps (cf.~\cite{CK11}). We tested the LSMC (cf.~\cite{LS01}) version of our JDOI algorithm. The results provide evidence of a strong variance reduction when compared with a simple application of the underlying LSMC algorithm and prove that applying the JDOI method on top of Monte Carlo based algorithms gives a powerful way to accelerate these methods. Finally, while we have only investigated LSMC based versions of our JDOI method, we are convinced that other JDOI versions could be introduced. In particular, our choice in favor of the LSMC algorithm was mainly due to its ubiquity in the financial industry and studying other JDOI versions, e.g.~GPR-JDOI or Deep-JDOI versions that either combine Gaussian process regression~(cf.~\cite{GMZ20}) or (deep) neural network based stopping-time algorithms (cf.~\cite{BCJ19}, \cite{BCJ20}) with our JDOI variance reduction technique, could be an interesting future avenue of research. \vspace{1em} \\

% appendices

\section*{Appendices}
\renewcommand{\theequation}{A.\arabic{equation}}
\subsection*{Appendix A: Derivations for Standard American Options}

This section provides complementary results for the derivatives in (\ref{Difference_InfGen}) as well as a derivation of the integral expression (\ref{Integral_StandardOption}) needed in the JDOI estimator of standard American options. A derivation of the respective quantities for American barrier options is presented in Appendix~B. \vspace{1em} \\
\noindent We first provide expressions for the derivatives in (\ref{Difference_InfGen}) and start by recalling that the (standard) Black~\&~Scholes Greeks are given by
\begin{align}
\partial_{S} V_{E}^{\bar{X},\mathcal{P}}(t,s_{0},\nu_{0},\eta_{0}; K) = -e^{-\delta(T-t)} \, \mathcal{N}\bigg( -d_1\bigg(\frac{s_0}{K},\bar{\sigma}_{T-t},T-t\bigg) \bigg), \hspace{5.5em} \\
 \partial_{\bar{\sigma}}V_{E}^{\bar{X},\mathcal{P}}(t,s_{0},\nu_{0},\eta_{0}; K) = K e^{-r(T-t)}\, \phi_{\mathcal{N}}\bigg(d_2\bigg(\frac{s_0}{K},\bar{\sigma}_{T-t},T-t\bigg)\bigg)\sqrt{T-t}, \hspace{4em} \\
\partial_{\bar{\sigma}}^{2} V_{E}^{\bar{X},\mathcal{P}}(t,s_{0},\nu_{0},\eta_{0}; K) = \frac{d_1\big(\frac{s_0}{K},\bar{\sigma}_{T-t},T-t\big) d_2\big(\frac{s_0}{K},\bar{\sigma}_{T-t},T-t\big)}{\bar{\sigma}_{T-t}} \,  \partial_{\bar{\sigma}}V_{E}^{\bar{X},\mathcal{P}}(t,s_{0},\nu_{0},\eta_{0}; K) , \\
 \partial_{S}\partial_{\bar{\sigma}}V_{E}^{\bar{X},\mathcal{P}}(t,s_{0},\nu_{0},\eta_{0}; K) = \frac{\partial_{\bar{\sigma}}V_{E}^{\bar{X},\mathcal{P}}(t,s_{0},\nu_{0},\eta_{0}; K)}{s_{0}} \left( 1- \frac{d_1\big(\frac{s_0}{K},\bar{\sigma}_{T-t},T-t\big)}{\bar{\sigma}_{T-t} \sqrt{T-t}} \right), \hspace{1.3em}
\end{align}
where we have used the notation introduced in~(\ref{Def_d1d2}), and $\phi_{\mathcal{N}}(\cdot)$ to denote the standard normal density. Then, using (\ref{Deter_Variance}) we get, for $\bullet \in \{\nu,\eta\}$, that

\begin{align}
\partial_{\bullet}[\bar{\sigma}_{T-t}]
= \frac{1}{2\bar{\sigma}_{T-t}}\partial_{\bullet}\left[ \bar{\sigma}_{T-t}^2 \right], \hspace{5.5em} \\
\partial_{\bullet}^{2}[\bar{\sigma}_{T-t}] = -\frac{1}{4\bar{\sigma}_{T-t}^3}\left(\partial_{\bullet}\left[ \bar{\sigma}_{T-t}^2 \right] \right)^2 + \frac{1}{2\bar{\sigma}_{T-t}}\partial_{\bullet}^{2}\left[ \bar{\sigma}_{T-t}^2\right], 
\end{align}
where
\begin{align}
\partial_{\nu}\left[ \bar{\sigma}_{T-t}^2 \right]&=\frac{c_1}{\kappa_1(T-t)} \left(1- e^{-\kappa_1 (T-t)}\right),\\
\partial_{\eta}\left[ \bar{\sigma}_{T-t}^2 \right]&= \frac{c_2}{\kappa_2 (T-t)}\frac{1-e^{-\kappa_2\theta_2 (T-t)}}{\eta_0 + (\theta_2 - \eta_0)e^{-\kappa_2\theta_2 (T-t)}}, \\
\partial_{\nu}^{2}\left[\bar{\sigma}_{T-t}^2\right]
&=0,\\
\partial_{\eta}^{2}\left[\bar{\sigma}_{T-t}^2\right]
&=-\frac{c_2}{\kappa_2 (T-t)}\left(\frac{1-e^{-\kappa_2\theta_2 (T-t)}}{ \eta_0 + \left( \theta_2-\eta_0 \right)e^{-\kappa_2\theta_2 (T-t)}}\right)^2,
\end{align}
\noindent and combining these expressions, we finally arrive at the following results for the Greeks in the GBS model:
\begin{align*}
\partial_{\bullet}V_{E}^{\bar{X},\mathcal{P}}(t,s_{0},\nu_{0},\eta_{0}; K)  = \partial_{\bar{\sigma}}V_{E}^{\bar{X},\mathcal{P}}(t,s_{0},\nu_{0},\eta_{0}; K) \, \partial_{\bullet}[\bar{\sigma}_{T-t}] , \hspace{6.3em} \\
\partial_{\bullet}^{2}V_{E}^{\bar{X},\mathcal{P}}(t,s_{0},\nu_{0},\eta_{0}; K)  = \partial_{\bar{\sigma}}^{2} V_{E}^{\bar{X},\mathcal{P}}(t,s_{0},\nu_{0},\eta_{0}; K) \big(  \partial_{\bullet}[\bar{\sigma}_{T-t}]  \big)^2 + \partial_{\bar{\sigma}} V_{E}^{\bar{X},\mathcal{P}}(t,s_{0},\nu_{0},\eta_{0}; K) \, \partial_{\bullet}^{2}[\bar{\sigma}_{T-t}], \\
\partial_{S} \partial_{\bullet}V_{E}^{\bar{X},\mathcal{P}}(t,s_{0},\nu_{0},\eta_{0}; K)  = \partial_{S}\partial_{\bar{\sigma}}V_{E}^{\bar{X},\mathcal{P}}(t,s_{0},\nu_{0},\eta_{0}; K) \,  \partial_{\bullet}[\bar{\sigma}_{T-t}] . \hspace{5em}
\end{align*}

\noindent We now turn to a derivation of the integral expression~(\ref{Integral_StandardOption}) and note that
\begin{align}
\label{App.B2:BS-type fomrula for V_E_bar(X)}
\int_\mathbb{R} & V_{E}^{\bar{X},\mathcal{P}}(t,s_0e^y,\nu_0,\eta_0; K) \varphi_Y^{\text{mix}}(y)\, dy \nonumber \\
 & \hspace{2.5em} =  \int_{\mathbb{R}} \left \{ Ke^{-r(T-t)} \mathcal{N}\bigg( - d_2\bigg(\frac{s_0}{K},\bar{\sigma}_{T-t},T-t\bigg) - \frac{y}{\bar{\sigma}_{T-t} \sqrt{T-t}} \bigg)\bigg)  \right.\nonumber \\
& \hspace{7.5em} \left. - s_0e^ye^{-\delta (T-t)} \mathcal{N} \bigg( -d_1\bigg(\frac{s_0}{K},\bar{\sigma}_{T-t},T-t\bigg) - \frac{y}{\bar{\sigma}_{T-t} \sqrt{T-t}} \bigg)\bigg) \right \} \varphi_Y^{\text{mix}}(y)\,  dy.
\end{align}
Therefore, we need to compute two integrals:
\begin{enumerate}
\item[$i)$] For the first part, we note that
\begin{align}
\int_{\mathbb{R}} & Ke^{-r(T-t)} \mathcal{N}\bigg( - d_2\bigg(\frac{s_0}{K},\bar{\sigma}_{T-t},T-t\bigg) - \frac{y}{\bar{\sigma}_{T-t} \sqrt{T-t}} \bigg)\bigg)\varphi_{Y}^{\text{mix}}(y)\, dy \hspace{15em} \nonumber \\
\nonumber & = Ke^{-r(T-t)} \Bigg \{ p_u \sum_{i=1}^m p_i \underbrace{\int_0^{\infty} \mathcal{N}\bigg(- d_2\bigg(\frac{s_0}{K},\bar{\sigma}_{T-t},T-t\bigg) - \frac{y}{\bar{\sigma}_{T-t} \sqrt{T-t}}\bigg) a_i e^{-a_i y}\, dy}_{=:I_{2,i}^{+}\left (\frac{s_0}{K},\bar{\sigma}_{T-t},T-t \right) } + \\
& \hspace{7.5em} +  q_d \sum_{j=1}^n q_j \underbrace{\int_{-\infty}^0  \mathcal{N} \bigg( - d_2\bigg(\frac{s_0}{K},\bar{\sigma}_{T-t},T-t\bigg) - \frac{y}{\bar{\sigma}_{T-t}\sqrt{T-t}}\bigg) \bigg) b_j e^{b_j y}\, dy}_{=:I_{2,j}^{-}\left (\frac{s_0}{K},\bar{\sigma}_{T-t},T-t \right) } \Bigg \},
\end{align}
\noindent and after a few manipulations -- integration by parts, grouping, and substitution -- we arrive at the following results:
\begin{align}
 I_{2,i}^{+}(\chi, \zeta, \xi)  = \mathcal{N}\Big( - d_{2}(\chi, \zeta, \xi) \Big)  -  e^{ d_{2}(\chi, \zeta, \xi) a_{i} \zeta \sqrt{\xi} + \frac{1}{2} a_{i}^{2} \zeta^{2} \xi } \, \mathcal{N} \Big( -d_{2}(\chi, \zeta, \xi)-a_{i} \zeta \sqrt{\xi} \Big), \\
 I_{2,j}^{-}(\chi, \zeta, \xi) = \mathcal{N}\Big( - d_{2}(\chi, \zeta, \xi) \Big)  +  e^{ -d_{2}(\chi, \zeta, \xi) b_{j} \zeta \sqrt{\xi} + \frac{1}{2} b_{j}^{2} \zeta^{2} \xi } \, \mathcal{N} \Big( d_{2}(\chi, \zeta, \xi)-b_{j} \zeta \sqrt{\xi} \Big). \hspace{0.3em} 
\end{align}
\item[$ii)$] Similarly, for the second part, we have that
\begin{align}
\int_{\mathbb{R}} & s_0e^y e^{-\delta (T-t)} \mathcal{N} \bigg( - d_1 \bigg(\frac{s_0}{K},\bar{\sigma}_{T-t},T-t\bigg) - \frac{y}{\bar{\sigma}_{T-t} \sqrt{T-t}} \bigg)\bigg) \varphi_{Y}^{\text{mix}}(y) \, dy \hspace{15em} \nonumber \\
\nonumber & = s_0e^{-\delta (T-t)} \Bigg \{ p_u \sum_{i=1}^m p_i \underbrace{\int_0^{\infty} \mathcal{N}\bigg(-d_1\bigg(\frac{s_0}{K},\bar{\sigma}_{T-t},T-t\bigg) - \frac{y}{\bar{\sigma}_{T-t} \sqrt{T-t}}\bigg)\bigg) a_i e^{-(a_i-1)y}\, dy}_{=:{} I_{1,i}^{+}\left (\frac{s_0}{K},\bar{\sigma}_{T-t},T-t \right)}  \\
& \hspace{6.5em} + q_d\sum_{j=1}^n q_j \underbrace{\int_{-\infty}^0 \mathcal{N}\bigg( -d_1\bigg(\frac{s_0}{K},\bar{\sigma}_{T-t},T-t\bigg) - \frac{y}{\bar{\sigma}_{T-t}\sqrt{T-t}} \bigg)\bigg) b_j e^{(b_j + 1)y}\, dy }_{=:{} I_{1,j}^{-}\left (\frac{s_0}{K},\bar{\sigma}_{T-t},T-t \right)} \Bigg \},
\end{align}
and computing $I_{1,i}^{+}(\cdot)$ and $I_{1,j}^{-}(\cdot)$ gives:
\begin{align}
 I_{1,i}^{+}(\chi, \zeta, \xi)  & = \frac{a_{i}}{a_{i} -1} \bigg \{ \mathcal{N}\Big( - d_{1}(\chi, \zeta, \xi) \Big)  \nonumber \\
  & \hspace{6em} - e^{ d_{1}(\chi, \zeta, \xi) (a_{i}-1) \zeta \sqrt{\xi} + \frac{1}{2} (a_{i}-1)^{2} \zeta^{2} \xi } \, \mathcal{N} \Big( -d_{1}(\chi, \zeta, \xi)-(a_{i}-1) \zeta \sqrt{\xi} \Big) \bigg \}, \\
 I_{1,j}^{-}(\chi, \zeta, \xi)  & = \frac{b_{j}}{b_{j} + 1} \bigg \{ \mathcal{N}\Big( - d_{1}(\chi, \zeta, \xi) \Big)  \nonumber \\
  & \hspace{6em} + e^{ -d_{1}(\chi, \zeta, \xi) (b_{j}+1) \zeta \sqrt{\xi} + \frac{1}{2} (b_{j}+1)^{2} \zeta^{2} \xi } \, \mathcal{N} \Big( d_{1}(\chi, \zeta, \xi)-(b_{j}+1) \zeta \sqrt{\xi} \Big) \bigg \},
\end{align}
\end{enumerate}
\noindent Consequently, we can combine the above results and arrive, upon setting
\begin{align}
\Psi_{2}(\chi, \zeta, \xi) & := p_{u} \sum \limits_{i=1}^{m} p_{i} \, I_{2,i}^{+}(\chi, \zeta, \xi) +  q_{u} \sum \limits_{j=1}^{n} q_{j} \, I_{2,j}^{-}(\chi, \zeta, \xi) , \label{PSI_1_first}\\
\Psi_{1}(\chi, \zeta, \xi) & := p_{u} \sum \limits_{i=1}^{m} p_{i} \, I_{1,i}^{+}(\chi, \zeta, \xi) +  q_{u} \sum \limits_{j=1}^{n} q_{j} \, I_{1,j}^{-}(\chi, \zeta, \xi), \label{PSI_1_second}
\end{align}
\noindent at the following (Black-Scholes like) representation of (\ref{Integral_StandardOption}):
\begin{align}
\hspace{1em} \int_\mathbb{R} V_{E}^{\bar{X},\mathcal{P}}(t,s_0e^y,\nu_0,\eta_0;K)&\varphi_Y^{\text{mix}}(y)dy \nonumber \\
&\hspace{-15mm} = Ke^{-r(T-t)} \Psi_2\bigg(\frac{s_0}{K},\bar{\sigma}_{T-t},T-t\bigg)- s_0e^{-\delta (T-t)}\Psi_1 \bigg(\frac{s_0}{K},\bar{\sigma}_{T-t},T-t\bigg),
\end{align}
This gives the result.
\subsection*{Appendix B: Derivations for American Barrier Options}
This section collects complementary results for the derivatives and integral expression, needed in (\ref{Difference_InfGen}) as part of the JDOI estimator for American barrier options. \vspace{1em} \\
To start, we note that analytical results for the Greeks of (regular) European up-and-out put options can be readily obtained by combining Relation (\ref{Def_UOP}) with the derivations provided for the (standard) European put in Appendix~A. Indeed, using Relation (\ref{Def_UOP}), we first arrive at the following expressions for the Delta, Vega(s), Vomma(s), and Vanna(s) of a European up-and-out put option
\begin{align}
\partial_{S} V_{E}^{\bar{X},\mathcal{UOP}}(t,s_{0},\nu_{0},\eta_{0}; K,H) = \partial_{S} V_{E}^{\bar{X},\mathcal{P}} (t,s_{0},\nu_{0},\eta_{0};K) \hspace{19.3em} \nonumber \\
 + \left( \frac{H}{s_{0}}\right)^{2(\gamma - \frac{1}{2})} \left\{ \frac{2(\gamma-1)}{H} V_{E}^{\bar{X},\mathcal{P}}\bigg(t,\frac{H^2}{s_{0}},\nu_{0},\eta_{0};K \bigg) + \frac{H}{s_{0}} \partial_{S} V_{E}^{\bar{X},\mathcal{P}}\bigg(t,\frac{H^2}{s_{0}},\nu_{0},\eta_{0};K \bigg) \right\}, \hspace{1em} \\
\partial_{\bullet} V_{E}^{\bar{X},\mathcal{UOP}}(t,s_{0},\nu_{0},\eta_{0}; K,H) = \partial_{\bullet} V_{E}^{\bar{X},\mathcal{P}} (t,s_{0},\nu_{0},\eta_{0};K) \hspace{21.4em} \nonumber \\
\hspace{1.5em} + \left( \frac{H}{s_{0}}\right)^{2(\gamma - 1)} \left \{ \frac{2(r-\delta)}{\bar{\sigma}_{T-t}^{4}} \partial_{\bullet} \left[ \bar{\sigma}_{T-t}^{2} \right] \log\bigg(\frac{H}{s_{0}}\bigg) V_{E}^{\bar{X},\mathcal{P}}\bigg(t,\frac{H^2}{s_{0}},\nu_{0},\eta_{0};K \bigg) - \partial_{\bullet} V_{E}^{\bar{X},\mathcal{P}}\bigg(t,\frac{H^2}{s_{0}},\nu_{0},\eta_{0};K \bigg) \right \},  \\
\partial_{\bullet}^{2} V_{E}^{\bar{X},\mathcal{UOP}}(t,s_{0},\nu_{0},\eta_{0}; K,H) = \partial_{\bullet}^{2} V_{E}^{\bar{X},\mathcal{P}} (t,s_{0},\nu_{0},\eta_{0};K) - \frac{4(r-\delta)}{\bar{\sigma}_{T-t}^{4}} \partial_{\bullet} \left[ \bar{\sigma}_{T-t}^{2} \right] \log\bigg(\frac{H}{s_{0}}\bigg) \left( \frac{H}{s_{0}}\right)^{2(\gamma - 1)}  \hspace{3em} \nonumber \\
\times \left \{ \left( \frac{(r-\delta)}{\bar{\sigma}_{T-t}^{4}} \partial_{\bullet} \left[ \bar{\sigma}_{T-t}^{2} \right] \log\bigg(\frac{H}{s_{0}}\bigg) + \frac{1}{\bar{\sigma}_{T-t}^{2}} \right)  V_{E}^{\bar{X},\mathcal{P}}\bigg(t,\frac{H^2}{s_{0}},\nu_{0},\eta_{0};K \bigg) - \partial_{\bullet} V_{E}^{\bar{X},\mathcal{P}}\bigg(t,\frac{H^2}{s_{0}},\nu_{0},\eta_{0};K \bigg) \right \}  \hspace{1.5em} \nonumber \\
\hspace{1.5em}  + \left( \frac{H}{s_{0}}\right)^{2(\gamma - 1)} \left \{ \frac{2(r-\delta)}{\bar{\sigma}_{T-t}^{4}} \partial_{\bullet}^{2} \left[ \bar{\sigma}_{T-t}^{2} \right] \log\bigg(\frac{H}{s_{0}}\bigg) V_{E}^{\bar{X},\mathcal{P}}\bigg(t,\frac{H^2}{s_{0}},\nu_{0},\eta_{0};K \bigg) - \partial_{\bullet}^{2} V_{E}^{\bar{X},\mathcal{P}}\bigg(t,\frac{H^2}{s_{0}},\nu_{0},\eta_{0};K \bigg) \right \},  \\
\partial_{S} \partial_{\bullet} V_{E}^{\bar{X},\mathcal{UOP}}(t,s_{0},\nu_{0},\eta_{0}; K,H) = \partial_{S} \partial_{\bullet} V_{E}^{\bar{X},\mathcal{P}} (t,s_{0},\nu_{0},\eta_{0};K) - \frac{2(r-\delta)}{\bar{\sigma}_{T-t}^{4}} \partial_{\bullet} \left[ \bar{\sigma}_{T-t}^{2} \right] \left( \frac{H}{s_{0}}\right)^{2(\gamma-\frac{1}{2})}  \hspace{3.9em} \nonumber \\
  \times \left \{ \left( \frac{2(\gamma -1)}{H} \log\bigg(\frac{H}{s_{0}}\bigg) + \frac{1}{H} \right) V_{E}^{\bar{X},\mathcal{P}}\bigg(t,\frac{H^2}{s_{0}},\nu_{0},\eta_{0};K \bigg) + \log\bigg(\frac{H}{s_{0}}\bigg) \frac{H}{s_{0}} \, \partial_{S} V_{E}^{\bar{X},\mathcal{P}}\bigg(t,\frac{H^2}{s_{0}},\nu_{0},\eta_{0};K \bigg) \right \}  \nonumber \\
+ \left( \frac{H}{s_{0}}\right)^{2(\gamma-\frac{1}{2})} \left \{ \frac{2(\gamma -1)}{H} \, \partial_{\bullet} V_{E}^{\bar{X},\mathcal{P}}\bigg(t,\frac{H^2}{s_{0}},\nu_{0},\eta_{0};K \bigg) + \frac{H}{s_{0}} \, \partial_{S} \partial_{\bullet} V_{E}^{\bar{X},\mathcal{P}}\bigg(t,\frac{H^2}{s_{0}},\nu_{0},\eta_{0};K \bigg) \right \} ,\hspace{1em} 
\end{align}
\noindent with $\bullet \in \{\nu,\eta \}$, and combining these results with the derivations of Appendix~A subsequently leads to the respective analytical expressions. \vspace{1em} \\
As a last step, we provide a derivation of Integral Term (II) in Equation~(\ref{UOP_Int_TERM}). Following the derivations in Appendix~A, we compute, for $i \in \{1, \ldots, m \}$, $j \in \{1, \ldots,n \}$, the following integrals
\begin{align}
 I_{2,i}^{+,\gamma}\bigg(\frac{H^{2}}{s_{0} K }, \bar{\sigma}_{T-t}, T-t \bigg) & := \int_{0}^{\infty} \mathcal{N} \bigg( - d_{2}\bigg(\frac{H^{2}}{s_{0} K }, \bar{\sigma}_{T-t}, T-t \bigg) + \frac{y}{\bar{\sigma}_{T-t} \sqrt{T-t}}\bigg) \, a_{i} \, e^{-\left(a_{i} + 2(\gamma -1 ) \right) y} \, dy , \label{UOP_Int2_Eq1} \\
I_{2,j}^{-,\gamma}\bigg(\frac{H^{2}}{s_{0} K }, \bar{\sigma}_{T-t}, T-t \bigg) & := \int_{-\infty}^{0} \mathcal{N} \bigg( - d_{2}\bigg(\frac{H^{2}}{s_{0} K }, \bar{\sigma}_{T-t}, T-t \bigg) + \frac{y}{\bar{\sigma}_{T-t} \sqrt{T-t}}\bigg) \, b_{j} \, e^{\left(b_{j} - 2(\gamma -1 ) \right) y} \, dy,\\
I_{1,i}^{+,\gamma}\bigg(\frac{H^{2}}{s_{0} K }, \bar{\sigma}_{T-t}, T-t \bigg) & := \int_{0}^{\infty} \mathcal{N} \bigg( - d_{1}\bigg(\frac{H^{2}}{s_{0} K }, \bar{\sigma}_{T-t}, T-t \bigg) + \frac{y}{\bar{\sigma}_{T-t} \sqrt{T-t}}\bigg) \, a_{i} \, e^{-\left(a_{i} + 2 \gamma -1 \right) y} \, dy,\\
I_{1,j}^{-,\gamma}\bigg(\frac{H^{2}}{s_{0} K }, \bar{\sigma}_{T-t}, T-t \bigg) & :=\int_{-\infty}^{0} \mathcal{N} \bigg( - d_{1}\bigg(\frac{H^{2}}{s_{0} K }, \bar{\sigma}_{T-t}, T-t \bigg) + \frac{y}{\bar{\sigma}_{T-t} \sqrt{T-t}}\bigg) \, b_{j} \, e^{\left(b_{j} - 2 \gamma + 1 \right) y} \, dy, \label{UOP_Int2_Eq4}
\end{align}
\noindent and obtain that
\begin{align}
 I_{2,i}^{+,\gamma}(\chi, \zeta, \xi) & = \frac{a_{i}}{a_{i} + 2(\gamma-1)} \bigg \{ \mathcal{N}\Big( - d_{2}(\chi, \zeta, \xi) \Big)  +  e^{ - d_{2}(\chi, \zeta, \xi) \left(a_{i} + 2(\gamma -1 )\right) \zeta \sqrt{\xi} + \frac{1}{2} (a_{i} + 2 (\gamma -1 ))^{2} \zeta^{2} \xi } \hspace{5em} \nonumber \\
 & \hspace{19em} \times  \mathcal{N} \Big( d_{2}(\chi, \zeta, \xi)-(a_{i} + 2(\gamma -1 )) \zeta \sqrt{\xi} \Big) \bigg \}, \\
 I_{2,j}^{-,\gamma}(\chi, \zeta, \xi) & = \frac{b_{j}}{b_{j} - 2(\gamma-1)} \bigg \{ \mathcal{N} \Big( - d_{2}(\chi, \zeta, \xi) \Big)  -  e^{d_{2}(\chi, \zeta, \xi) \left(b_{j} - 2(\gamma -1 )\right) \zeta \sqrt{\xi} + \frac{1}{2} (b_{j} - 2 (\gamma -1 ))^{2} \zeta^{2} \xi } \hspace{6em} \nonumber \\
 & \hspace{19em} \times  \mathcal{N} \Big( -d_{2}(\chi, \zeta, \xi)-(b_{j} - 2(\gamma -1 )) \zeta \sqrt{\xi} \Big) \bigg \}, \\
 I_{1,i}^{+,\gamma}(\chi, \zeta, \xi) & = \frac{a_{i}}{a_{i} + 2 \gamma -1 } \bigg \{ \mathcal{N} \Big( - d_{1}(\chi, \zeta, \xi) \Big)  +  e^{ - d_{1}(\chi, \zeta, \xi) \left(a_{i} + 2 \gamma -1 \right) \zeta \sqrt{\xi} + \frac{1}{2} (a_{i} + 2 \gamma -1 )^{2} \zeta^{2} \xi }  \hspace{6em} \nonumber \\
 & \hspace{19em} \times  \mathcal{N} \Big( d_{1}(\chi, \zeta, \xi)-(a_{i} + 2 \gamma -1 ) \zeta \sqrt{\xi} \Big) \bigg \}, \\
 I_{1,j}^{-,\gamma}(\chi, \zeta, \xi) & = \frac{b_{j}}{b_{j} - 2 \gamma + 1} \bigg \{ \mathcal{N} \Big( - d_{1}(\chi, \zeta, \xi) \Big)  -  e^{d_{1}(\chi, \zeta, \xi) \left(b_{j} - 2 \gamma + 1\right) \zeta \sqrt{\xi} + \frac{1}{2} (b_{j} - 2 \gamma + 1)^{2} \zeta^{2} \xi } \hspace{6em} \nonumber \\
 & \hspace{19em} \times  \mathcal{N} \Big( -d_{1}(\chi, \zeta, \xi)-(b_{j} - 2 \gamma + 1) \zeta \sqrt{\xi} \Big) \bigg \} .
\end{align} 
\noindent Here, it is important to note that the (integrability) conditions imposed in (\ref{UOP_Int2_Eq1})-(\ref{UOP_Int2_Eq4}), 
\begin{align}
a_{i} + 2(\gamma -1) > 0, \hspace{1em} \forall i \in \{1, \ldots, m \}, \hspace{1.75em} \mbox{and} \hspace{1.75em} b_{j} - 2 \gamma + 1 > 0, \hspace{1em} \forall j \in \{1, \ldots, n \},  \label{Int_COND_STRONG}
\end{align}
\noindent are not directly required to hold under general mixed-exponential jump-diffusion dynamics, however these will be naturally satisfied in practice (cf.~\cite{LV20}, \cite{FMV20}). \vspace{1em} \\
\noindent Combining the above results and upon setting
\begin{align}
\Psi_{B,2}(\chi, \zeta, \xi, \gamma) & := p_{u} \sum \limits_{i=1}^{m} p_{i} \, I_{2,i}^{+,\gamma}(\chi, \zeta, \xi) +  q_{u} \sum \limits_{j=1}^{n} q_{j} \, I_{2,j}^{-,\gamma}(\chi, \zeta, \xi) , \label{PSI_2_first} \\
\Psi_{B,1}(\chi, \zeta, \xi, \gamma) & := p_{u} \sum \limits_{i=1}^{m} p_{i} \, I_{1,i}^{+,\gamma}(\chi, \zeta, \xi) +  q_{u} \sum \limits_{j=1}^{n} q_{j} \, I_{1,j}^{-,\gamma}(\chi, \zeta, \xi), \label{PSI_2_second}
\end{align}
\noindent we finally arrive at the following representation of Integral (II) in (\ref{UOP_Int_TERM}):
\begin{align}
\int_{\mathbb{R}} & V_{E}^{\bar{X},\mathcal{P}}\bigg(t,\frac{H^2}{s_{0}e^{y}},\nu_{0},\eta_{0};K \bigg) \,e^{-2(\gamma - 1)y} \, \varphi_{Y}^{mix}(y) \, dy \nonumber \\
& \hspace{2.5em} = K e^{-r(T-t)} \Psi_{B,2}\bigg(\frac{H^{2}}{s_{0} K }, \bar{\sigma}_{T-t}, T-t , \gamma \bigg) - \frac{H^2}{s_{0}} e^{-\delta (T-t)} \Psi_{B,1}\bigg(\frac{H^{2}}{s_{0} K }, \bar{\sigma}_{T-t}, T-t , \gamma \bigg).
\end{align}
This gives the result.

% bibliography
\printbibliography

\end{document}